\documentclass{aastex63}

\usepackage{amsmath}
\usepackage{mathtools}

\usepackage{xcolor}

\usepackage{rotating}
\usepackage{tikz}

\usepackage{comment}

\received{April 26, 2022}
\revised{\today}
\submitjournal{ApJ}

\shorttitle{Numerical Investigation of the Scales of Complexity and Coherence within ICMEs}
\shortauthors{Scolini et al.}

\begin{document}

\title{Characteristic Scales of Complexity and Coherence within Interplanetary Coronal Mass Ejections: 
Insights from Spacecraft Swarms in Global Heliospheric Simulations}

\author[0000-0002-5681-0526]{Camilla Scolini}
\affiliation{Institute for the Study of Earth, Oceans, and Space, University of New Hampshire, Durham, NH, USA}
\affiliation{Cooperative Programs for the Advancement of Earth System Science (CPAESS), University Corporation for Atmospheric Research, Boulder, CO, USA}
\correspondingauthor{Camilla Scolini}
\email{camilla.scolini@unh.edu}

\author[0000-0002-9276-9487]{R\'eka M. Winslow}
\affiliation{Institute for the Study of Earth, Oceans, and Space, University of New Hampshire, Durham, NH, USA}

\author[0000-0002-1890-6156]{No\'e Lugaz}
\affiliation{Institute for the Study of Earth, Oceans, and Space, University of New Hampshire, Durham, NH, USA}

\author[0000-0002-1743-0651]{Stefaan Poedts}
\affiliation{Department of Mathematics/Centre for mathematical Plasma Astrophyics, KU Leuven, Belgium}
\affiliation{Institute of Physics, University of Maria Curie-Skłodowska, Lublin, Poland}



\begin{abstract} 
Many aspects of the three-dimensional (3-D) structure and evolution of {i}nterplanetary {c}oronal {m}ass {e}jections (ICMEs) remain unexplained. 
Here, we investigate two main topics: (1) the coherence scale of magnetic fields inside ICMEs, and (2) the dynamic nature of ICME magnetic complexity.
We simulate ICMEs interacting with different solar winds using the linear force-free spheromak model incorporated into the EUHFORIA model. We place a swarm of $\sim20000$ spacecraft in the 3-D simulation domain and characterize ICME magnetic complexity and coherence at each spacecraft based on simulated time series. 
Our simulations suggest that ICMEs retain a lower complexity and higher coherence along their magnetic axis, but that a characterization of their global complexity requires crossings along both the axial and perpendicular directions. 
For an ICME of initial half {angular} width of $45^\circ$ that does not interact with other large-scale solar wind structures, global complexity can be characterized by as little as 7--12 spacecraft separated by $25^\circ$, but the minimum number of spacecraft rises to 50--65 (separated by $10^\circ$) if interactions occur.
Without interactions, ICME coherence extends for $45^\circ$, $20^\circ$--$30^\circ$, $15^\circ$--$30^\circ$, and $0^\circ$--$10^\circ$ for $B$, $B_\phi$, $B_\theta$, and $B_r$, respectively. Coherence is also lower in the ICME west flank compared to the east flank due to Parker spiral effects. 
Moreover, coherence is reduced by a factor of 3--6 by interactions with solar wind structures. 
Our findings help constrain some {of the} critical scales {that control} the evolution of ICMEs and {aid in} the planning of future dedicated multi-spacecraft missions.
\end{abstract}



\section{Introduction} 
\label{sec:introduction}

Coronal mass ejections {\citep[CMEs; see e.g.][]{Webb2012}} are large eruptions of plasma and magnetic field from the Sun into interplanetary space. {When probed in situ by spacecraft crossing their structures,} they are referred to as interplanetary CMEs {\citep[ICMEs; see e.g.][]{Bothmer1998, Cane2003, Kilpua2017}. ICMEs are among the largest and most energetic transients in the heliosphere}, where they shape the space environment and can drive severe space weather disturbances at Earth and other planets {\citep[see e.g.][]{Gosling1990, Tsurutani1992, Zhang2007, Kilpua2017b, Lee2017, Slavin2019, Winslow2020}}.
{ICMEs are} characterized by a variety of {in-situ} signatures, with their core component being the so-called magnetic ejecta \citep[ME;][]{Winslow2015}, i.e.\ a region of enhanced magnetic field corresponding to the original structure erupted at the Sun, which may or may not be preceded by a forward shock and a turbulent sheath formed during propagation \citep{Kilpua2017}. {The entire propagating structure, composed of the ME and its associated shock and sheath (if present) is referred to as the ICME. We use the term ME as a more general term than magnetic cloud, which follows the strict definition of \citet{Burlaga1981}.}
{Typically, MEs are associated with a low plasma $\beta$ ($\ll 1$) near 1~au, indicating that the magnetic pressure dominates over the thermal pressure and that their internal dynamics and 3-D structure are controlled primarily by magnetic forces. MEs also often exhibit rotations in the magnetic field components, which are} compatible with a configuration of magnetic field bundles twisting around a central axis {\citep{Burlaga1988, Lepping1990, Kilpua2017}}. 
This has led to the proliferation of flux-rope (FR) models to describe {the ME} local and global magnetic structure {\citep[see, e.g.,][]{Lepping1990, Mulligan2001, Marubashi2007, Hidalgo2012, Hidalgo2013, Isavnin2016, NievesChinchilla2016, NievesChinchilla2018b, Weiss2021}. FR models have also been successful in describing the CME pre-eruptive and eruptive} configuration, early evolution in the solar corona, and interplanetary evolution up to Mars and beyond {\citep[e.g.][]{Vourlidas2014, Patsourakos2020, Palmerio2021, Davies2022}.}

ICMEs evolve significantly as they propagate through the solar wind. In particular, interactions with other large-scale interplanetary structures such as high speed streams \citep[HSSs;][]{Cranmer2017}, stream interaction regions \citep[SIRs;][]{Richardson2018}, or other ICMEs can induce {dramatic} changes {to ICME structures} \citep{Manchester2017, Lugaz2017}. {Therefore, u}nderstanding the effect of different propagation scenarios on ICMEs is necessary to characterize their {evolution across a wide range of spatial and temporal scales. To this end, recent studies have explored the changes in magnetic complexity (i.e., the degree of similarity/deviation from an ideal FR configuration) that MEs can undergo during propagation. In-depth investigations of {MEs} observed by radially-aligned spacecraft provided ample evidence that {MEs evolve and} tend to increase their magnetic complexity during propagation through interplanetary space, primarily due to interaction with other large-scale solar wind structures {\citep{Mulligan1999, NievesChinchilla2012, Good2016, Winslow2016, Winslow2021a, Winslow2021b}}. \citet{Scolini2022} provided further statistical support to these results through an investigation of 31 MEs observed in radial alignment between Mercury and 1~au. Furthermore, \citet{Scolini2021b} conducted a numerical investigation of the probability of detecting a complexity increase between an inner and outer spacecraft in perfect radial alignment, finding that the interaction with HSSs and SIRs doubles the probability for an {MEs} to undergo complexity changes. To date it has not been possible to completely characterize the spatial (i.e.\ longitudinal and latitudinal) magnetic complexity distribution of MEs and its evolution with heliocentric distance due to the limited number of spacecraft crossing individual {ME} structures.}

Direct probing of {ICMEs} relies on in situ spacecraft measurements obtained at a given location in space (i.e.\ along a 1-D trajectory). {However, single-point measurements only provide information on the local properties of an ICME at a particular stage of its evolution, and it remains unclear to what extent they are indicative of their global structure \citep[e.g.][]{Moestl2012, Al-Haddad2013, Al-Haddad2018, Marubashi2015, Lugaz2018}}.
Multi-point observations {at the same heliocentric distance but} at different locations in space {represent} the best opportunity to probe the 3-D structure of ICMEs {at a given stage of their evolution, as they allow for a direct probing of the in situ ICME characteristics at multiple locations while mitigating the effect of local variations \citep[e.g.][]{Moestl2012, Weiss2021}. However, these remain relatively scarce, as the majority of ICMEs are still probed in situ by single spacecraft, and the maximum number of spacecraft crossings at a given heliocentric distance per individual event is limited to three \citep[e.g.][]{Burlaga1981, Mulligan2001, Kilpua2011, Lugaz2022}}. Additionally, such measurements have {so far been} obtained only near 1~au, either by multiple spacecraft in L1 orbits (at small {angular} separations $\le 1^\circ$), or for spacecraft angular separations above $10^\circ$, with other heliocentric distances and critical scales around $1^\circ-10^\circ$ remaining unprobed \citep{Lugaz2018}.

Prominent open questions related to the 3-D nature of ICMEs are whether {their MEs} are magnetically coherent objects, and at what spatial scales such {a} coherence exists. Strictly speaking, magnetic coherence within an {ME} refers to its ability to respond to external perturbations in a coherent (i.e.\ solid-like) manner \citep{Owens2017}. {Coherent behavior requires information to propagate across an ME structure, and magnetic forces to be sufficiently strong to resist external deformation forces \citep{Owens2017}}. Following an analytical approach based on the consideration of expansion and Alfv\'en speeds within {MEs} at 1~au, \citet{Owens2017} and \citet{Owens2020} estimated an upper limit for the angular scale of magnetic coherence to $\sim 26^\circ$, which is significantly smaller than the typical angular width of {MEs} \citep[around $40^\circ-60^\circ$; e.g.][]{Yashiro2004, Kilpua2011, Good2016}.
From an observational standpoint, the correlation of magnetic field components among different locations within an {ME} has been regarded as a proxy for magnetic coherence {\citep{Matsui2002, Farrugia2005, Lugaz2018, Ala-Lahti2020}}, although it remains unclear how the ability of information to propagate across an {ME} affects its local properties, and how this relates to the ability of the same {ME} to use such information in order to resist deformation by external factors and behave as a coherent structure \citep{Owens2020}. The investigation by \citet{Lugaz2018} across various angular separations pointed towards the potential existence of two characteristic scales of correlation within MEs near 1~au: one related to the magnetic field components (around $4^\circ–7^\circ$) and one for the total magnetic field (around $14^\circ–20^\circ$). Yet, limitations on the data available prevented the exploration of the whole parameter space, particularly {at scales} between $1^\circ$ and $10^\circ$. 
 
It remains extremely challenging to answer some of the most fundamental questions on the nature of {MEs}, particularly with respect to their magnetic complexity and coherence and their evolution with heliocentric distance. 
{In this paper, we address these topics by turning to numerical simulations and exploiting} their 3-D capabilities to investigate the following science questions:
(1) What is the spatial distribution of magnetic complexity within {MEs}?
(2) Across what spatial scales could {MEs} behave as magnetically coherent objects?
(3) How does all of the above depend on the heliocentric distance and the specific {propagation} history of an {ME}, such as its interaction with other large-scale solar wind structures?
%
We address the aforementioned questions by means of global heliospheric simulations performed with the EUropean Heliospheric FORecasting Information Asset \citep[EUHFORIA;][]{Pomoell2018}, using idealized solar wind set-up as in \citet{Scolini2021b} and simulating {CMEs/MEs} using a linear force-free spheromak model \citep{Verbeke2019}. 
{T}his work represents the first-ever attempt to specifically address the problem of magnetic coherence {by analyzing numerical simulation results using techniques typically employed in the analysis of in situ observations}.

{The} paper is structured as follows.
In Section~\ref{sec:methods}, we introduce the numerical model and the methods used to investigate the magnetic complexity and coherence of {MEs} using global heliospheric simulations. 
In Section~\ref{sec:results_euhforia_spheromak}, the analysis of two simulations considering different interaction scenarios between an {ME} and the surrounding solar wind is presented, and results are discussed with respect to baseline values {obtained for a  stand-alone ideal spheromak magnetic structure,} and to the propagation history of each {ME}.
In Section~\ref{sec:discussion}, we examine possible implications of our results for future dedicated multi-spacecraft missions orbiting close to the ecliptic plane. Finally, in Section~\ref{sec:conclusions}, we summarize our conclusions and outlook. 

\section{Methods}
\label{sec:methods}

\subsection{Numerical Set-Up}

In this work, we build on the analysis of the numerical simulations previously presented by \citet{Scolini2021b}, that were performed using the EUHFORIA 3-D magnetohydrodynamic (MHD) model of the inner heliosphere \citep{Pomoell2018}.
To evaluate the interaction of ICMEs with different solar wind structures, 
two background solar wind configurations are {simulated}: 
the first one (hereafter ``run~A'') includes a uniform solar wind of speed $450$~km~s$^{-1}$ except for a low-inclination heliospheric current/plasma sheet (HCS/HPS).
The second one (hereafter ``run~B'') differs from the one above by having a more highly inclined HCS/HPS, and by the inclusion of a HSS of radial speed equal to $675$~km~s$^{-1}$ and circular cross-section of half {angular} width equal to $30^\circ$.
In both cases, the HPS meridional profile is parametrized using the description in \citet{Odstrcil1996}, resulting in a solar wind speed as low as 300~km~s$^{-1}$ near the HCS. 
This choice has been made to ensure full control over the ICME propagation and interaction with solar wind structures, and the comparability between different runs.
In the two runs, {CMEs} are {initialized} as linear force-free spheromak {structures as} described by \cite{Verbeke2019}, {with} the following set of geometric, kinematic, and magnetic parameters:
radial speed equal to $v_R = 800$~km~s$^{-1}$;\
initial half {angular} width of $\omega/2 = 45^\circ$;\
positive chirality {(${H=+1}$)} with axial tilt ${\gamma}=90^\circ$ with respect to the northward direction \citep[corresponding to a SWN flux-rope type; see][]{Bothmer1998};\
toroidal magnetic flux {(${\varphi_t}$)} equal to $10^{14}$~Wb (corresponding to a magnetic field strength of $\sim 25$~nT at 1~au).\
Because of the pressure imbalance between the CME and the surrounding solar wind upon insertion in the heliospheric domain \citep[leading to an expansion of the CME structure, as shown by][]{Scolini2019, Scolini2021a}, 
the effective initial CME front speed is $\sim 1100$~km~s$^{-1}$, which results in a fast CME that drives an interplanetary shock and sheath.
Such a combination of initial parameters is representative of those of a typical fast CME with a reconnected flux of the order of $10^{14}$~Wb \citep[][]{Pal2018}.
The CME initial direction is chosen to reproduce two end-member scenarios of interaction with different solar wind structures:
in run~A, the CME is inserted across the HCS/HPS at $(\theta, \phi) = (0^\circ, 0^\circ)$;
in run~B, the CME is inserted across the HSS at $(\theta, \phi) = (5^\circ, 0^\circ)$, in a configuration similar to that of CMEs originated from ``anemone'' active regions \citep[e.g.][]{Lugaz2011, Sharma2020}. The CME insertion time is arbitrarily set on January 1, 2020 at 00:00~UT.\
A detailed discussion of the model set-up is available in \citet{Scolini2021b}.

\begin{figure*}
\centering
{\includegraphics[width=\hsize]{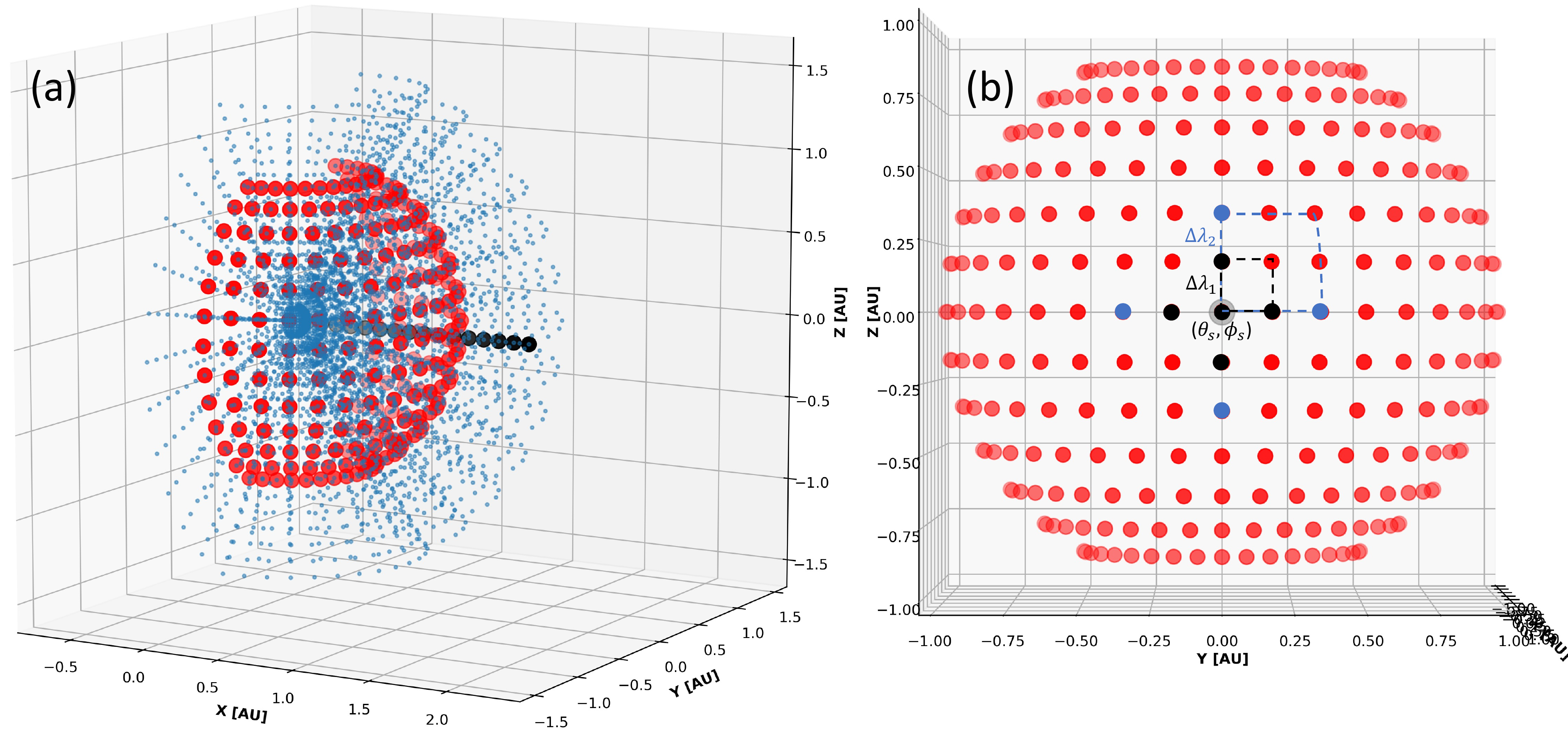}}
\caption{
{Spatial distribution of the spacecraft swarm used in EUHFORIA simulations.}
(a): Angled view showing the global distribution of virtual spacecraft {in 3-D space (in blue). Red dots indicate the} virtual spacecraft located at 1~au. {Black} dots mark the virtual spacecraft located along the {CME} initial direction.
(b): Front view showing only the virtual spacecraft at 1~au. For a reference spacecraft serving as seed location $(\theta_s, \phi_s)$, the re-binning of ME types with nearby spacecraft for two different $\Delta \lambda$ values is indicated by the dashed polygons. For the same reference spacecraft, first neighbors considered under the rook criterion employed to determine the Moran's $I$ coefficient, for two different $\Delta \lambda$, are indicated as black and blue colored dots. 
} 
\label{fig:app_sketch_spacecraft_positions_01}
\end{figure*}

Our simulation domain expands from 0.1 to 2.0~au in the radial direction ($r$), covering $\pm 80^\circ$ in the latitudinal direction ($\theta$), and $\pm 180^\circ$ in longitude ($\phi$), with a uniform grid composed of $512(r) \times 80 (\theta) \times 180 (\phi)$ cells. In our simulation domain, we place a set of virtual spacecraft where the time-dependent MHD parameters are extracted in a similar way to that done by real spacecraft monitoring the in situ plasma conditions in space.
Virtual spacecraft are placed at longitudinal and latitudinal separations of $\Delta \lambda = 5^\circ$, spanning $\pm 90^\circ$ in longitude from the {CME} initial direction, and covering the full range of latitudes in the simulation domain. They are also uniformly distributed in the radial direction between 0.11 and 1.61~au (i.e., from the model inner boundary to the orbit of Mars) with a 0.1~au radial separation. 
Overall, a swarm of 18944 virtual spacecraft (1184 per heliocentric distance) is placed in the model domain in each simulation.
{Each virtual spacecraft records time profiles of the standard MHD output in EUHFORIA: the velocity vector, magnetic field vector, density, and thermal pressure in Heliocentric Earth Equatorial (HEEQ) coordinates.}
Figure~\ref{fig:app_sketch_spacecraft_positions_01}~(a) provides a global view of the {locations of the virtual spacecraft considered, while} a detailed discussion about their use in this paper's analysis is provided in the following sections.

\subsection{Metrics of Magnetic Complexity}
\label{subsec:metrics_complexity}

Time-dependent information on the MHD variables {(i.e.\ velocity vector, magnetic field vector, density, and thermal pressure in HEEQ coordinates)} at each simulated spacecraft are used to investigate the spatial and temporal evolution of the magnetic complexity characterizing the simulated ICMEs, and particularly their ME structures, as described below.

\subsubsection{Frequency of Magnetic Ejecta Types}
\label{subsubsec:me_types}

At each virtual spacecraft located at coordinates $(r, \theta, \phi)$ in the simulation domain, we identify whether an ME signature is present {in the simulated time series} using the identification scheme provided in \citet{Scolini2021b}. {This classification scheme considers the magnetic field and plasma $\beta$ to identify the presence or absence of an ME following a preceding interplanetary shock, and to determine its front and rear boundaries. The preceding shock is identified based on jump conditions on the radial speed, density, and magnetic field.} If an ME is detected, we characterize its complexity {based on the maximum rotation detected in the magnetic field components} by using the classification scheme previously introduced by \citet{Scolini2021b} and based on previous observational works by \citet{NievesChinchilla2018, NievesChinchilla2019}. {Our classification differs from the one by \citet{NievesChinchilla2018, NievesChinchilla2019} in the fact that it is adapted to reflect the full range of rotation angles expected to be observed when crossing a spheromak structure, ranging between $0^\circ$ and $360^\circ$ as detailed by \citet{Scolini2021b}.} In summary, each ME signature is classified as $F_{270}$, $F_{180}$, $F_{90}$, $F_{30}$ or $E$ when at least one component (i.e.\ $B_R$, $B_T$, or $B_N$) rotates by $\ge 270^\circ$, $\ge 180^\circ$, $\ge 90^\circ$, $\ge 30^\circ$, or $< 30^\circ$, respectively.
Additionally, we assign a numerical index, defined as ``complexity index'' ($\mathbb{C}$), to each ME class detected in situ, in order to rank the level of complexity of the detected structure: $\mathbb{C}=0$ for $F_{270}$ class, $\mathbb{C}=1$ for $F_{180}$ class, $\mathbb{C}=2$ for $F_{90}$ class, $\mathbb{C}=3$ for $F_{30}$ class, and $\mathbb{C}=4$ for $E$ class.
As a result, a complexity scale between 0 and 4 is defined, with higher complexity values indicating {higher deviations from the FR signatures expected for an ideal spheromak, which exhibit rotations larger than $270^\circ$ for most crossings near the spheromak central axis \citep[see Figure~6 in][]{Scolini2021b}}. Additionally, a value of $\mathbb{C}=5$ is assigned to shock-only ICME signatures, while $\mathbb{C}=6$ is assigned to all spacecraft locations that did not detect any ICME-related signature. {Example signatures for each ME class are provided in the appendix of \citet{Scolini2021b}. We note this definition of complexity is solely based on the magnetic field signatures associated with a given {ME}, and it does not take into account any plasma information nor the properties of its preceding shock and sheath. This choice was made to be consistent with observational works {in the past $\sim15$ years, which could} only make use of magnetic field data at distances other than 1~au \citep[][]{Winslow2016, Winslow2021a, Scolini2022, Davies2022}. However, it will be important for future investigations of ICME complexity to asses them holistically, i.e., by looking at changes in all aspects of the ICME as opposed to just the magnetic parameters within the ME, to understand complexity changes from a more fundamental standpoint.}

{In our simulation runs A and B,} the frequency of ME types detected at different spacecraft is then investigated as a function of the heliocentric distance, and of the angular separation $\Delta \lambda$ of the simulated spacecraft crossing through the {ME} structure. 
To explore different spatial scales, we resample the initial spacecraft distribution (uniformly separated in longitude and latitude by $\Delta \lambda = 5^\circ$) to progressively larger $\Delta \lambda${, covering $\Delta \lambda$ between $5^\circ$ and $45^\circ$}. After defining a ``seed'' location $(\theta_s, \phi_s)$ corresponding to the reference location for the resampling of the spacecraft distribution, spacecraft located between $(\theta_s + n \Delta \lambda, \phi_s + m \Delta \lambda)$ and $(\theta_s + (n+1) \Delta \lambda, \phi_s + (m+1) \Delta \lambda)$ are binned into the $n$-th ($m$-th) bin of the new spatial distribution in latitude (longitude). A representation of such a re-binning process is provided in Figure~\ref{fig:app_sketch_spacecraft_positions_01}~(b). Information from individual spacecraft binned within the same $\Delta A = \Delta \lambda \times \Delta \lambda$ areal unit is then combined by taking the mode of the ME type distribution across $\Delta A$, which corresponds to the most probable ME class that would be observed by a spacecraft crossing anywhere within $\Delta A$.

Because results expressed in terms of $\Delta \lambda$ also depend on the specific CME half {angular} width ($\omega/2$) prescribed at insertion through the domain inner boundary, we also express angular separations as fractions of the CME initial half {angular} width ($F_{\omega/2}${, ranging between 0 and 1}), which are independent of the prescribed CME half {angular} width and can therefore be extrapolated to CMEs of different sizes. When relevant, the associated number of virtual spacecraft crossing the {ME} structure is also reported.

In the following, we consider the occurrence of ME types detected at different spacecraft to provide a global measure of {ME} magnetic complexity: the lowest the $\mathbb{C}$ values detected within an ME, the lowest its global complexity. 
Furthermore, we identify the characteristic scale of magnetic complexity as the maximum angular separation $\Delta \lambda$ at which spacecraft are able to detect the same two most frequent ME classes detected by spacecraft with minimum $\Delta \lambda = 5^\circ$ separation, at all heliocentric distances. We  clarify this choice in the identification of the characteristic scale of magnetic {complexity in Sections~\ref{subsec:ideal_spheromak_structure} and~\ref{subsec:euhforia_me_types}, when discussing the results for an ideal spheromak structure and for a spheromak propagating through a quiet solar wind.}

\subsubsection{Spatial Autocorrelation of Magnetic Ejecta Types: the Moran’s $I$}
\label{subsubsec:moransi}

To evaluate how ordered is the spatial distribution of {ME} complexity at each heliocentric distance $r$ considered, we use the Moran's $I$ \citep{Moran1950} as a measure to quantify the spatial autocorrelation {(i.e.\ the presence of systematic spatial variation in a certain variable)} of different ME types detected within the simulated ICMEs. 
The Moran's $I$ at a given heliocentric distance $r$ is defined as
\begin{equation}
    I(r) = \frac{N}{W} \frac{\sum^N_{i=1} \sum^N_{j=1} w_{ij} (x_i-\bar{x})(x_j-\bar{x})}{\sum^N_{k=1} (x_k-\bar{x})^2}{,}
    \label{eqn:moransI}
\end{equation}
where $N$ is the total number of spacecraft observations at a given heliocentric distance $r$;
$x_i=\mathbb{C}_i$ is the value of the ME complexity index at a given location $i$ corresponding to coordinates $(r, \theta_i, \phi_i)$;
$x_j=\mathbb{C}_j$ is the value of the ME complexity index at a location $j$ corresponding to coordinates $(r, \theta_j, \phi_j)$;
and $\bar{x}=\bar{\mathbb{C}}$ is the mean of all $x_i=\mathbb{C}_i$ values, for $i=1,...,N$.
The contiguity between two spatial units is assessed using the rook criterion, which takes as neighbors any pair of spacecraft that are aligned in longitude or latitude (i.e.\ whose location cells on a grid share an edge). For the specific spacecraft swarm configuration used in this work's simulations, the neighbors of a given spacecraft using the rook criterion are indicated by the blue and black colored dots in Figure~\ref{fig:app_sketch_spacecraft_positions_01}~(b) for two different $\Delta \lambda$ values. $w_{ij}$ is the weight of location $i$ relative to $j$, and it assumes a value of 1 for pairs of spacecraft that are aligned in longitude or latitude, and a value of 0 for pairs of spacecraft that are diagonally aligned. $W$ is the sum of all weights $w_{ij}$ over the set of $N$ locations considered.
Under such a prescription, a Moran's $I(r)$ close to 1 indicates a well-ordered, highly-clustered spatial distribution of ME types and is taken as indication of the lowest complexity state possible for an {ME} magnetic structure; a Moran's $I(r)$ approaching 0 indicates a random arrangement of ME types in space, and therefore indicates the highest complexity state. In the following, we consider $I> 0.5$ as indicative of a {low-complexity,} highly clustered spatial distribution of ME classes within an {ME}, and $I\le0.5$ as indicative of a {high-complexity,} randomly-arranged distribution.

{Similarly to the occurrence of ME types, we compute} the Moran's $I$ for different spacecraft separations $\Delta \lambda$, and {we} evaluate its changes as a function of the heliocentric distance to assess the role of interactions between ICMEs and solar wind structures in randomizing the internal magnetic field of {MEs}. 
The Moran's $I$ for different spatial scales is obtained from the resampled spatial distribution of the complexity index $\mathbb{C}$, as described in Section~\ref{subsubsec:me_types}.
In addition to the occurrence of ME classes discussed in Section~\ref{subsubsec:me_types}, {we use the Moran's $I$ as} a complementary global metric to quantify the degree of complexity within an {ME}.

\subsubsection{Spatial Autocorrelation of Magnetic Ejecta Types: the Local Moran's $I$}
\label{subsubsec:local_moransi}

To gain information on the local properties of {ME} complexity, we {further inspect the spatial distribution of magnetic complexity through} the Local Moran's $I$ \citep[also known as Anselin's Local Indicator of Spatial Association;][]{Anselin1995}. This metric corresponds to the local formulation of the Moran's $I${, which is} calculated locally for each areal unit in the data spatial distribution. {Specifically}, the Local Moran's $I$ at a given location $(r, \theta_i, \phi_i)$ is defined as 
\begin{equation}
    I_i(r) =  (x_i-\bar{x}) \sum^{N_i}_{j=1} w_{ij} (x_j-\bar{x}){,}
    \label{eqn:localmoransI}
\end{equation}
where
$N_i$ is the number of neighboring spacecraft observations to the reference location $i$ corresponding to coordinates $(r, \theta_i, \phi_i)$;
$x_i=\mathbb{C}_i$ is the ME complexity index at location $i$;
$x_j=\mathbb{C}_j$ is the ME complexity index at location $j$ corresponding to coordinates $(r, \theta_j, \phi_j)$ neighboring location $i$;
and $\bar{x}=\bar{\mathbb{C}}$ is the mean of all $x_j=\mathbb{C}_j$ neighboring values of location $i$, for $j=1,...,N_i$.
As for Equation~\ref{eqn:moransI}, $w_{ij}$ is the weight of location $i$ relative to $j$, as determined by the rook criterion.

With respect to the specific problem considered in this work, the Local Moran's $I$ proves particularly useful to investigate the spatial distribution of {ME} complexity as a function of the heliocentric distance, and to quantify to what degree interactions with interplanetary structures affect the magnetic complexity in different regions of an {ME}. In particular, to locate the source of complexity changes, we track the $(\theta, \phi)$ location of the maximum of the Local Moran's $I$ ($\mathrm{max}(I_i)$) with heliocentric distance, and relate it to the conditions of the surrounding solar wind along the corresponding propagation direction.


\subsection{Metrics of Magnetic Coherence}
\label{subsec:metrics_coherence}

As previously done for magnetic complexity, we consider time-dependent information on the MHD plasma variables at each simulated spacecraft to investigate the spatial and temporal evolution of the magnetic coherence characterizing the simulated ICMEs, and particularly their ME structures, as described below. Our final aim is to characterize the spatial properties and scale of {ME} magnetic coherence and its evolution depending on the characteristics of the solar wind encountered by an ICME along its journey through the heliosphere. 

\subsubsection{Scale of Magnetic Coherence: Correlation of Magnetic Field Profiles at Different Spacecraft}

\begin{figure*}
\centering
{\includegraphics[width=\hsize]{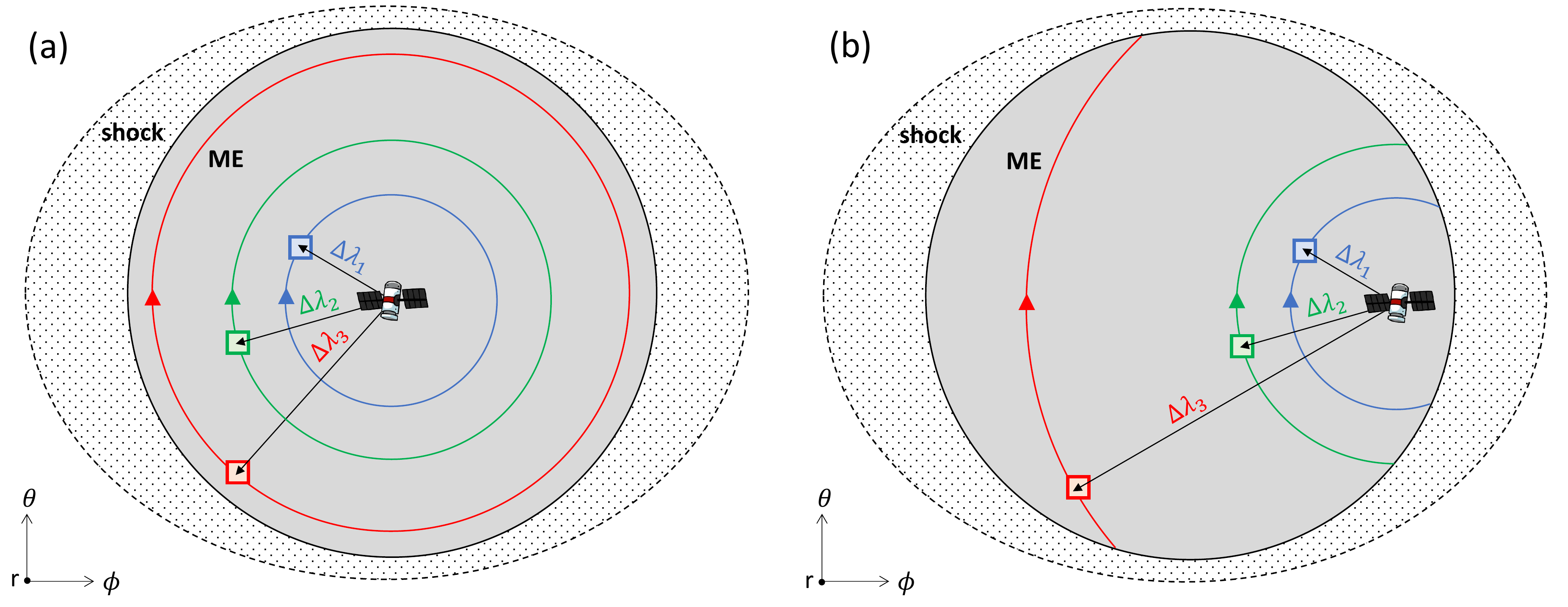}}
\caption{
{Sketch of the methodology used to determine the average correlation scale of the magnetic field profiles around a reference observer crossing through the ME center {(a)} and west {(b)} region. Different angular scales ($\Delta \lambda_i$) are indicated in different colors. The ICME is plotted as propagating heads on towards the observer.}
} 
\label{fig:app_sketch_spacecraft_positions_02}
\end{figure*}

In the following, we investigate {ME} magnetic coherence in a similar manner as previous observational investigations, i.e.\ by quantifying the spatial scales at which the time series of the components and magnitude of the magnetic field within simulated {MEs} exhibit high levels of correlation. 
To achieve our goal, for each heliocentric distance $r$, we assess the correlation of the magnetic field profile measured within the ME at each spacecraft location $(r, \theta_i, \phi_i)$ with respect to selected reference positions $(r, \theta_\mathrm{ref}, \phi_\mathrm{ref})$. 
We then compute average correlations across all spacecraft located at a given angular separation from the reference spacecraft ($\Delta \lambda$), i.e.\ by taking the average along all directions around the reference location, for different angular separations.
By comparing the average correlations obtained at different angular separations, we determine the maximum angular separation at which two spacecraft are likely to observe correlated time profiles for a given magnetic field component. As correlation scales around a given reference spacecraft may be anisotropic along different directions, the correlation scale for a given magnetic field component is estimated by averaging across different directions around the reference observer and chosen as the maximum angular separation at which correlation coefficients are $\ge 0.6$ on average. Cartoons describing its derivation by averaging across different directions with respect to two different reference observers, {one crossing near the center of the ME and one near its west flank, for three given angular separations $\Delta \lambda$, are provided in Figure~\ref{fig:app_sketch_spacecraft_positions_02}~(a) and (b), respectively}. 
We note that our definition of correlation scale is different from those used by \citet{Lugaz2018} and \citet{Wicks2010} to investigate the ME and solar wind correlation lengths from in situ measurements. A justification on the different definition chosen in this study compared to previous studies on spatial correlations is provided in Section~\ref{subsec:euhforia_magnetic_coherence}, when discussing the results {for spheromaks propagating through different solar winds with respect to the baseline model consisting of an ideal spheromak structure}. 

\section{Results for Spheromak Structures Propagating in the Heliosphere}
\label{sec:results_euhforia_spheromak}

{In this section, we explore the evolution of magnetic complexity and coherence within {MEs}} propagating through different solar winds in EUHFORIA, and we discuss how deviations from the patterns retrieved for an ideal spheromak structure depend on the characteristics of the interaction with the surrounding solar wind. In light of the results by \citet{Scolini2021a, Scolini2021b}, we focus our discussion on heliocentric distances between 0.3~au and 1.6~au and neglect results between 0.1~au and 0.3~au, where ICME over-expansion and challenges in the determination of the ME boundaries from simulated time series in the early stage of the ICME propagation might have otherwise influenced our results and conclusions.
{As a caveat to the interpretation of the results, we note that the chosen numerical set-up ensures more control over the various propagation effects affecting different ICME regions during propagation compared to performing simulations of real ICME events propagating through a magnetogram-extrapolated solar wind. The insertion of a perfectly-known global {CME} magnetic structure at the inner boundary of the heliospheric domain (located at 0.1~au in our simulations) also ensures enhanced control over the evolution of its magnetic structure, facilitating our interpretation of propagation effects. However, it is important to acknowledge that such an approach will likely result in a more regular/smooth magnetic configuration in simulated ICMEs compared to real ICMEs at the same heliocentric distance. To avoid drawing inappropriate conclusions, in the following we put care into the interpretation of EUHFORIA results with respect to a baseline model (i.e.\ a stand-alone ideal spheromak model).}

\subsection{Baseline Results: Complexity and Coherence in an Ideal Spheromak Structure}
\label{subsec:ideal_spheromak_structure}

{We start discussing the results of our investigation by summarizing the baseline values obtained from the consideration of an ideal, stand alone spheromak structure. Characterizing the baseline level of complexity and coherence in such an idealized condition represents a first but necessary step to later be able to conduct a meaningful interpretation of the results obtained for spheromak structures propagating through different solar wind conditions.}

{As a first global measure of ME magnetic complexity considered, we report the frequency of detection of different ME classes as a function of the angular separation of the spacecraft crossings in the ideal spheromak. 
We find that for spacecraft separations of $\Delta \lambda = 5^\circ$, about 80\% of the detections are $F_{270}$ (corresponding to $\mathbb{C}=0$) and $F_{30}$ (corresponding to $\mathbb{C}=3$), which therefore contribute the most to the overall complexity of the ME. This result justifies our choice of considering only the two most frequently detected ME types as metrics of global ME complexity, and it applies also to simulated spheromaks propagating through different ambient solar winds, as discussed in Section~\ref{subsec:euhforia_me_types}. On the other hand, less represented ME classes (i.e.\ $F_{180}$, $F_{90}$, and $E$ types, corresponding to $\mathbb{C}=1,2$, and $4$, respectively) are observed by less than 10\% of the spacecraft and quickly disappear as soon as the spacecraft angular separation becomes $\Delta \lambda \sim 15^\circ$ ($F_{\omega/2} \sim 0.33$) or larger. 
For increasing spacecraft angular separations, we find that $\Delta \lambda = 25^\circ$ (corresponding to $F_{\omega/2} \sim 0.55$ of the CME half angular width) is the maximum separation that guarantees a characterization of the magnetic complexity of the ideal spheromak as good as a $\Delta \lambda = 5^\circ$ separation in terms of the two most frequently observed ME classes (i.e.\ $F_{270}$ and $F_{30}$ types).}

{The second global measure of ME magnetic complexity considered in this work is the Moran's $I$. The Moran's $I$ is around 0.95 for angular separations $\Delta \lambda = 5^\circ$ ($F_{\omega/2}= 0.11$), and it decreases at larger angular scales. 
The decrease of $I$ for increasing $\Delta \lambda$ reflects the uncorrelated nature of ME signatures detected by spacecraft crossing the {ME} trajectories far apart. Overall, a spheromak structure exhibits a highly clustered spatial distribution of ME classes ($I\ge 0.5$, indicating a low complexity) when observed at scales $\Delta \lambda \lesssim 25^\circ$ (corresponding to $F_{\omega/2} \lesssim 0.55$), and a more randomly-arranged ME type distribution ($I < 0.5$, indicating a high complexity) at scales $\Delta \lambda = 25^\circ-45^\circ$ (corresponding to $F_{\omega/2} = 0.55-1$).} 

{Finally, as a proxy for the coherence of an ME, we consider the average correlation scales of the magnetic field profiles around three reference observers crossing near the east, center, and west portion of the ME structure.
We find that $B$, $B_\theta$ and $B_\phi$ show high correlations throughout the spheromak structure for all the observers considered, while the spatial correlation pattern for $B_r$ appears highly variable depending on the reference location considered, resulting in much smaller correlation scales ($\Delta \lambda \sim 0^\circ-25^\circ$ depending on the observer). 
While the exact characteristics of such a pattern are specific to the spheromak model considered in this study, a lower correlation in $B_r$ than in the remaining magnetic field components has also been reported in actual observations \citep{Lugaz2018} and by simple geometrical considerations, is expected to affect other flux rope models as well for observers crossing along different trajectories.}

\subsection{Characteristic Scales of Magnetic Complexity during Propagation}
\label{subsec:euhforia_me_types}

\begin{figure*}
\centering
{\includegraphics[width=0.7\hsize]{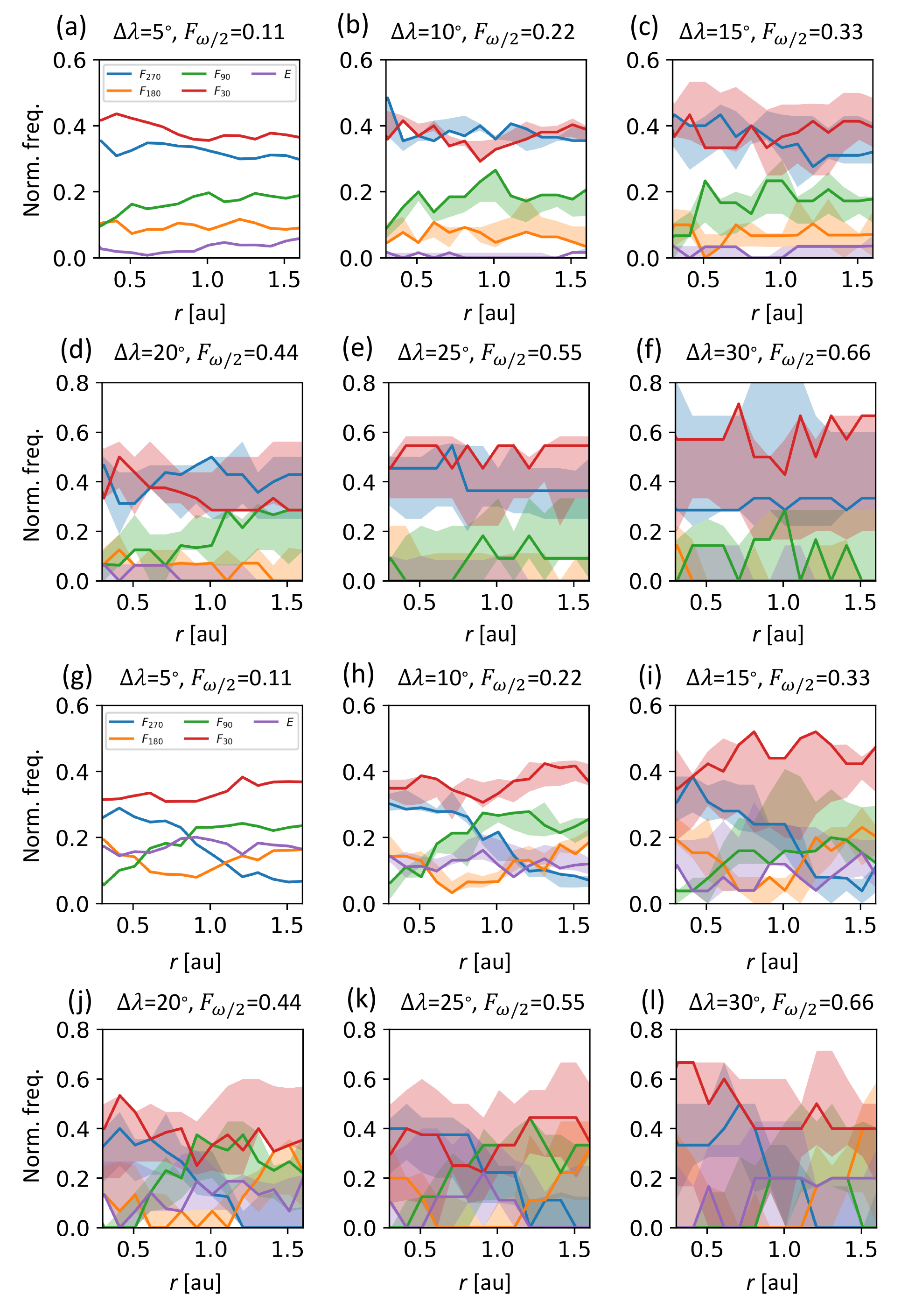}}
\caption{
Normalized frequency of ME types detected as a function of the heliocentric distance {and of the spacecraft angular separation}, for the two CMEs simulated in EUHFORIA.
(a)--(f): results for run~A.
(g)--(l): results for run~B.
Different panels show the results for different spacecraft separation considered.
{In all panels, the continuous lines indicate the results for a seed location at $(\theta_s, \phi_s)=(0^\circ, 0^\circ)$, while the shaded areas} indicate the uncertainty (i.e.\ the spread{, corresponding to the range of maximum and minimum values}) as obtained by picking 20 random seed locations.
} 
\label{fig:euhforia_ME_types}
\end{figure*}

{In what follows, we discuss the simulation results for EUHFORIA runs A and B with regards to scales of magnetic complexity as the ICME evolves during propagation.} 
Figure~\ref{fig:euhforia_ME_types} shows the {frequency of detection of different} ME classes recovered by virtual spacecraft crossing {the {MEs} through different trajectories} in EUHFORIA runs~A and B{, as a function of the heliocentric distance.} 
Figure~\ref{fig:euhforia_ME_types}~(a) shows that {for an} ICME propagating through a quiet, quasi-uniform solar wind (run~A) and observed by spacecraft at $5^\circ$ separation, {about 70\% of all detections are $F_{270}$ or $F_{30}$ ME classes (corresponding to $\mathbb{C}=0$ and $3$, respectively), which therefore contribute the most to the overall complexity of the {ME}}. This remains true for all heliocentric distances sampled. {This result applies not only to a spheromak propagating through a quasi-uniform solar wind, but also to a stand-alone ideal spheromak structure (see Section~\ref{subsec:ideal_spheromak_structure}), and it justifies our choice of considering only the two most frequently detected ME types as metrics of global {ME} complexity. Based on this result, we consider a {given} angular scale $\Delta \lambda$ (or the corresponding $F_{\omega/2}$) to provide a good representation of the global magnetic complexity of an {ME} structure when the two ME classes most frequently detected are well represented for all heliocentric distances.} 

The remaining three classes, i.e.\ $F_{180}$, $F_{90}$ and $E$ {(corresponding to $\mathbb{C}=1,2$ and $4$, respectively)}, are more rarely detected (each between 0\% and 20\%, depending on the heliocentric distance). This ranking is well represented up to $\Delta \lambda = 25^\circ$ {(Figure~\ref{fig:euhforia_ME_types}~(b)--(e))}, while at larger separations, the ranking between most and least frequently observed classes is {significantly altered depending on the seed location chosen} {(Figure~\ref{fig:euhforia_ME_types} (f))}.
We {therefore} conclude that the maximum $\Delta \lambda$ guaranteeing a good representation of the {ME} global magnetic complexity and its evolution with heliocentric distance is about $25^\circ$ for the case without interactions with other structures. This corresponds to about 7 to 12 observing spacecraft, and to spatial scales equal to $F_{\omega/2}=0.55$ {of the initial CME half angular width}. Such scales, as well as the occurrence of different ME types detected within the {ME}, are both comparable to the results retrieved from the analysis of an ideal spheromak structure ({as discussed in Section~\ref{subsec:ideal_spheromak_structure}}), indicating {that} the magnetic complexity of an {ME within an ICME} propagating through a quasi-uniform solar wind remains approximately constant during propagation, and at all times similar to the one of a pristine structure.

{In contrast, the distribution of ME classes is highly variable with heliocentric distance when interactions occur.}
Figure~\ref{fig:euhforia_ME_types}~(g) shows that in the case of an ICME interacting with a preceding HSS/SIR (run~B) and observed by spacecraft at $5^\circ$ separation, {results are similar to run~A only up until 0.9~au, with} the two dominant ME classes {being $F_{270}$ ($\mathbb{C}=0$) and $F_{30}$ ($\mathbb{C}=3$)}. At larger heliocentric distances, the second most frequently detected ME type changes from $F_{270}$ to the more complex $F_{90}$ type ($\mathbb{C}=2$){. This indicates that the {ME} increases its global complexity during propagation as a consequence of the interaction with the HSS/SIR \citep[consistent with previous findings by e.g.][]{Winslow2016, Winslow2021a,Scolini2021b, Scolini2022}.
Additionally, the ranking between most and least frequently observed classes} is well represented only up to spacecraft separations of $\Delta \lambda = 10^\circ$ {(Figure~\ref{fig:euhforia_ME_types}~(h))}, while at larger separations {it varies depending on the seed location chosen (Figure~\ref{fig:euhforia_ME_types}~(i) to (l))}.
We conclude that the distribution of ME types, and therefore of {ME} complexity as a whole, requires observations by about 50 to 65 spacecraft, separated by about $F_{\omega/2}=0.22$ {of the initial CME half angular width}. In this case, the characteristic scale necessary to capture {ME} complexity is less than half of the scale retrieved from the analysis of an ideal spheromak structure ({see Section~\ref{subsec:ideal_spheromak_structure}}), and {of an ME} not undergoing any interactions. This highlights that smaller angular separations must be used when investigating magnetic complexity within an {ME} that interacts with other large-scale structures compared to ICMEs that propagate through a quasi-uniform, quiet solar wind.

\subsection{{Spatial Autocorrelation of Magnetic Complexity:} Evolution of the Moran's $I$ during Propagation}
\label{subsec:euhforia_moransI}

{Next, we investigate the spatial distribution of magnetic complexity within the modeled MEs by considering the Moran's $I$ evolution with heliocentric distance.}
\begin{figure*}
\centering
{\includegraphics[width=0.8\hsize]{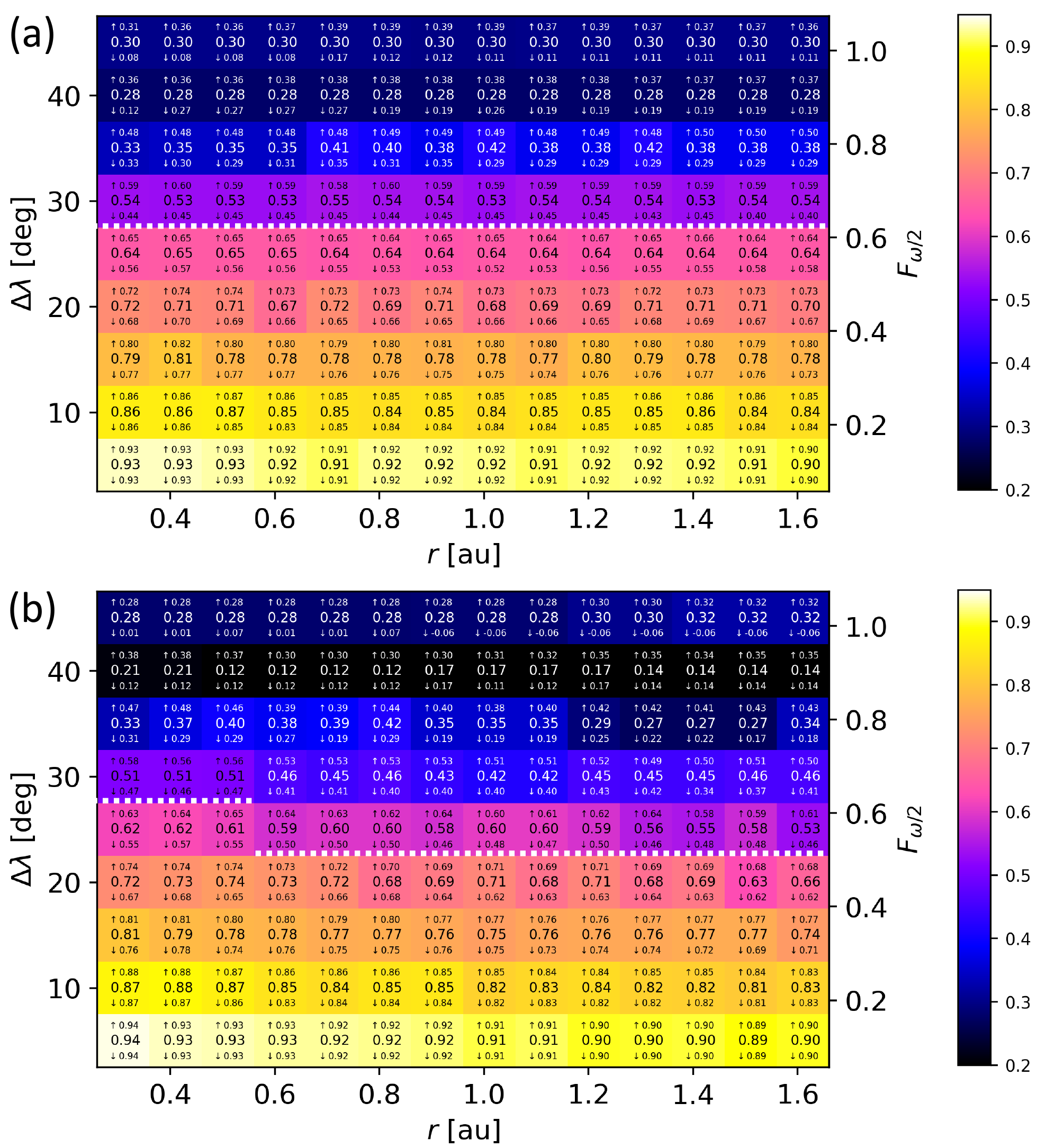}}
\caption{
Moran's $I$ as a function of the spacecraft separation $\Delta \lambda$ and heliocentric distance $r$, for the two ICMEs simulated in EUHFORIA.
(a): {Moran's $I$ resulting from} run~A. 
(b): {Moran's $I$ resulting from} run~B. 
The {color code and} values overplotted in each cell indicate the {Moran's $I$ obtained for a seed location at $(\theta_s, \phi_s)=(0^\circ, 0^\circ)$, while the $\uparrow$ and $\downarrow$ symbols indicate the} uncertainty (i.e.\ the spread{, corresponding to the range of maximum and minimum values}) as obtained by picking 20 random seed locations.
{The dotted white lines distinguish between regions where $I > 0.5$ (considered regions with well-ordered distribution of ME types and low magnetic complexity) and regions where $I \le 0.5$ (considered regions with randomly-ordered distribution of ME types and high magnetic complexity).}} 
\label{fig:euhforia_moransI}
\end{figure*}
%
{In run~A (non-interacting ICME), the Moran’s $I$ decreases with increasing $\Delta \lambda$ but is essentially unchanged during propagation towards larger heliocentric distances (Figure~\ref{fig:euhforia_moransI}~(a)). At 0.3~au, the range covered by the Moran's $I$ lies between $0.93^{+0.00}_{-0.00}$ {for $\Delta \lambda=5^\circ$} and $\sim 0.30^{+0.01}_{-0.22}$ {for $\Delta \lambda=45^\circ$}, which is very similar to values at 1.6~au (between $0.90^{+0.00}_{-0.00}$ {for $\Delta \lambda=5^\circ$} and $\sim 0.30^{+0.06}_{-0.19}$ {for $\Delta \lambda=45^\circ$}). Here, error bars indicate the uncertainty arising from the re-binning process after picking 20 random seed locations, and it therefore only affect angular scales larger than $5^\circ$.
Also, the $I$ values obtained for} $\Delta \lambda = 5^\circ$ are extremely similar to the values retrieved for an ideal spheromak structure observed at the same angular scale {(as discussed in Section~\ref{subsec:ideal_spheromak_structure})}. The maximum separation at which the Moran’s $I$ remains above 0.5 for all distances and seed locations considered is $\Delta \lambda = 25^\circ$ (corresponding to $F_{\omega/2}=0.55$ {of the initial CME half angular width}), as marked by the {white dotted} line in Figure~\ref{fig:euhforia_moransI}~(a). 
This {also indicates similar} levels of clustering in the spatial arrangement of the ME class between the {ME} in run~A and the ideal spheromak structure.
{From these results we conclude that s}pacecraft swarms composed of at least 7--12 evenly-spaced spacecraft crossing the same {ME} structure at different impact angles from the center to the flanks measure a highly-clustered distribution of ME types and generally a low level of {ME} magnetic complexity, while swarms that are more widely separated observe a high-complexity structure with a randomly-arranged ME type distribution.

The spatial autocorrelation of ME classes within an {ME} behaves differently when interactions with other structures affect its propagation.
In run~B (Figure~\ref{fig:euhforia_moransI}~(b)), {the {ME} initially has a similar complexity than in run~A, as indicated by the Moran’s $I$ ranging between $0.94^{+0.00}_{-0.00}$ for $\Delta \lambda=5^\circ$ and $0.28^{+0.00}_{-0.27}$ for $\Delta \lambda=45^\circ$} at 0.3~au. Small separations also report higher Moran's $I$ values than large separations: at this stage, {we observe} a low complexity ($I>0.5$) up to $\Delta \lambda = 25^\circ$ of separation (corresponding to $F_{\omega/2}=0.55$ {of the initial CME half angular width}), {and a high complexity} at larger separations. However, in contrast with run~A, {the characteristic scale of transition from a low ($I>0.5$) to a high ($I \le 0.5$) complexity decreases with heliocentric distance}, likely due to the interaction of the ICME with the HSS/SIR.
{Specifically, we observe a transition in the location of the $I=0.5$ threshold from $\Delta \lambda = 25^\circ$ (corresponding to $F_{\omega/2}=0.55$ of the initial CME half angular width) to $\Delta \lambda = 20^\circ$ of separation (corresponding to $F_{\omega/2} = 0.44$) near 0.6~au. By the time the ME reaches 1.6~au}, the range covered by the Moran's $I$ is between $0.90^{+0.00}_{-0.00}$ {for $\Delta \lambda=5^\circ$} and $\sim 0.32^{+0.00}_{-0.38}$ {for $\Delta \lambda=45^\circ$}, i.e.\ similar to the values retrieved in run~A, {but we observe a clear decreasing} trend with respect to the heliocentric distance at all scales $\lesssim 20^\circ$, where the Moran's $I$ values at 1.6~au are significantly smaller than the initial values retrieved at 0.3~au. {This can be seen in the fact that} the error bars inferred from the consideration of 20 random seed locations do not overlap. 
No clear trend with respect to the heliocentric distance is apparent at larger scales (where the values at 0.3~{au} and 1.6~au are consistent within the error bars {inferred from the consideration of 20 random seed locations}), indicating at these scales no changes {to} the global {ME} magnetic complexity are detected during propagation. 
Overall, these results indicate that the magnetic complexity quantified in terms of the spatial autocorrelation of ME types observed across different directions is affected by the interaction with the HSS/SIR structure {only at separation scales as large as} $\Delta \lambda = 20^\circ$ {(at spatial scales $F_{\omega/2} \le 0.44$ of the initial CME half angular width). This corresponds to having a minimum of 13--16 spacecraft crossings through the same {ME} to capture the evolutionary trends in the clustering of ME types, as} swarms separated by larger separations would consistently observe a highly-randomized spatial distribution of ME types {and complexity} regardless of the specific {ME propagation} history.

\begin{figure*}
\centering
{\includegraphics[width=\hsize]{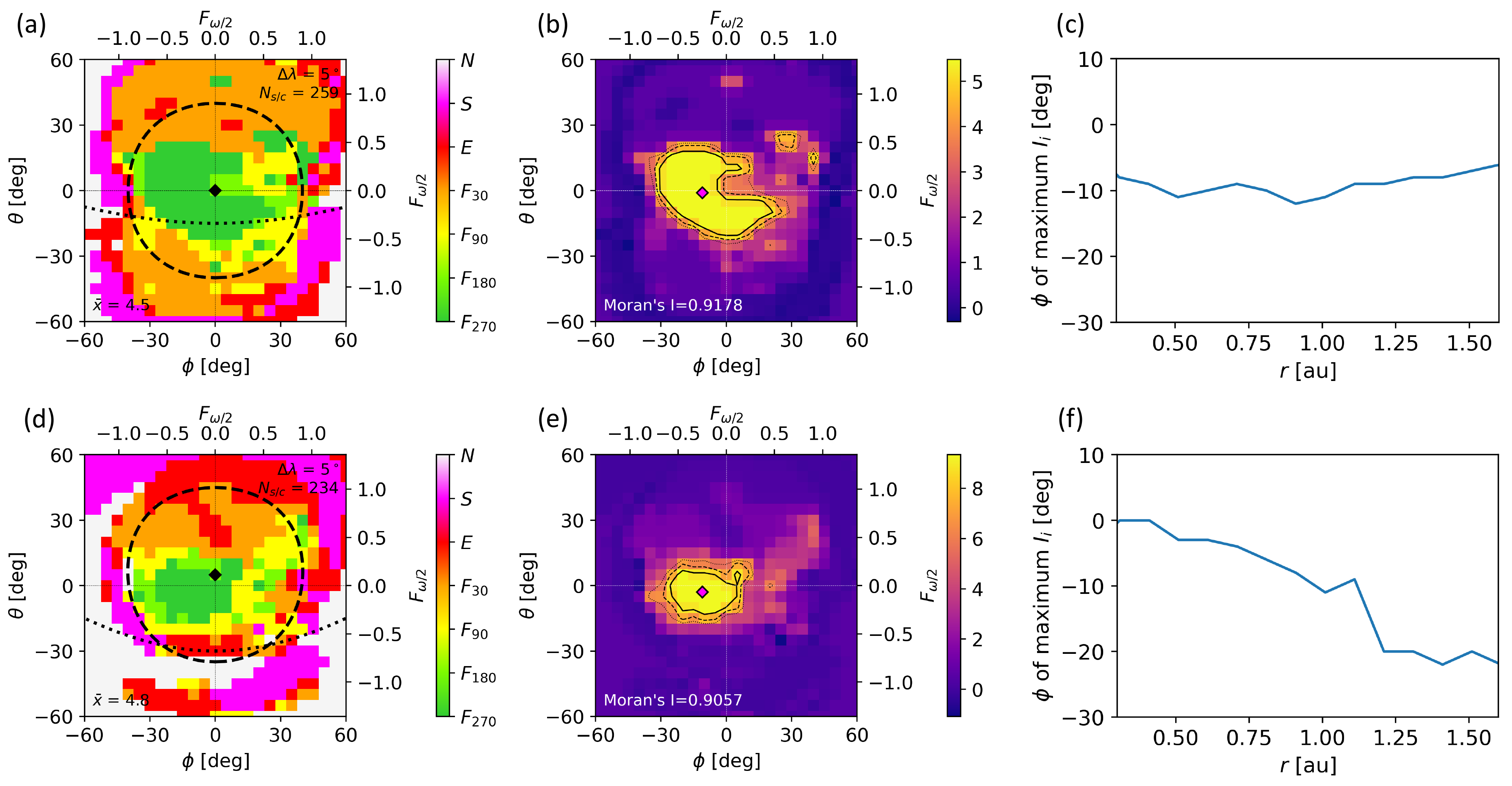}}
\caption{
Analysis of the Local Moran's $I$ {near 1~au} for the two ICMEs simulated in EUHFORIA.
(a), (d): ME type distribution at 1~au {(color coded)}, for run A and B, respectively.
(b), (e): distribution of the Local Moran's $I$ ($I_i$) at 1~au {(color coded)}, for run A and B, respectively. The locations of the maximum of the Local Moran's $I$ ($\mathrm{max}(I_i)$) are indicated by the magenta diamonds, while contours denote the $[0.9, 0.75, 0.6] \,\, \mathrm{max}(I_i)$ levels.
(c), (f): longitude of the maximum of the Local Moran's $I$, as a function of the heliocentric distance.
} 
\label{fig:euhforia_local_moransI}
\end{figure*}

To confirm the role of the interaction with the HSS/SIR as source of the Moran's $I$ increase with heliocentric distance in run~B, we consider the spatial distribution of the Local Moran's $I$ as a function of the heliocentric distance in both EUHFORIA runs.
{Figures~\ref{fig:euhforia_local_moransI}~(a) and (d) show the spatial distribution of ME types for runs A and B near 1~au which we use to derive the Local Moran's $I$ distributions reported in Figures~\ref{fig:euhforia_local_moransI}~(b) and (e).
To determine the complexity increase affecting different ME regions, we track the $(\theta, \phi)$ position of its maximum value (marking the lowest complexity region) as a function of the heliocentric distance.
{In run~A we find that the $\phi_{\mathrm{max}(Ii)}$ remains between $-5^\circ$ and $-10^\circ$ at all heliocentric distances considered (Figure~\ref{fig:euhforia_local_moransI}~(c)).} The fact that the location of lowest complexity within the ME does not change with heliocentric distance is consistent with the lack of interactions altering the ME structure.
{In run~B, on the other hand, we observe that the maximum Local Moran's $I$ progressively drifts towards the east with respect to the Sun--Earth line, with $\phi_{\mathrm{max}(Ii)}$ shifting from $0^\circ$ at 0.3~au to $-23^\circ$ at 1.6~au (Figure~\ref{fig:euhforia_local_moransI}~(f)).} In other words, during its propagation the ME retains a low complexity only on the easternmost region, while the westernmost region  progressively increases its complexity. This is consistent with the westernmost ME portion being more heavily affected by the interaction with the HSS/SIR, which may be the prime source of increased complexity in that region of the ME. 
Interestingly, we also note that even in run~A,  $\phi_{\mathrm{max}(Ii)}$ is slightly offset with} respect to the Sun--Earth line. In the absence of any interplanetary structure in the simulated ambient solar wind that could have induced such a longitudinal asymmetry in the spatial distribution of {ME} complexity, we consider this result to be intrinsic to the propagation of the ICME through a Parker magnetic field spiral, which presents an intrinsic asymmetry with respect to the ICME propagation direction. Such an interpretation is further corroborated by the results obtained from the spatial analysis of magnetic coherence presented below in Section~\ref{subsec:euhforia_magnetic_coherence}.

\subsection{Characteristic Scales of Magnetic Coherence during Propagation}
\label{subsec:euhforia_magnetic_coherence}

{In this section, we investigate the scale at which an ME may be able to respond to external perturbations in a coherent manner. For this purpose, we consider the correlation of magnetic field components between different observers as a proxy for the scale of magnetic coherence, similarly to previous observational investigations \citep{Wicks2010, Lugaz2018, Ala-Lahti2020}. Determining across which spatial scales MEs exhibit coherent characteristics is a first, necessary step towards a better understanding the global response of ME structures to external perturbations induced by the surrounding environment.
As reference directions to investigate the scales of magnetic coherence, we consider hypothetical observers crossing the {ME} structures in regions differently affected by the interaction of the ICME with the surrounding wind and, in run~B, with {the} large-scale structures therein. Such reference directions are located along the equatorial plane of the {ME} ($\theta_\mathrm{ref}=\theta_\mathrm{CME}$) and at $\phi_\mathrm{ref}$ angles of $-30^\circ$, $0^\circ$, and $30^\circ$, i.e.\ corresponding to observers crossing the {ME} magnetic structure on the east, center, and west regions near its central magnetic field axis.}
\begin{figure*}
\centering
{\includegraphics[width=\hsize]{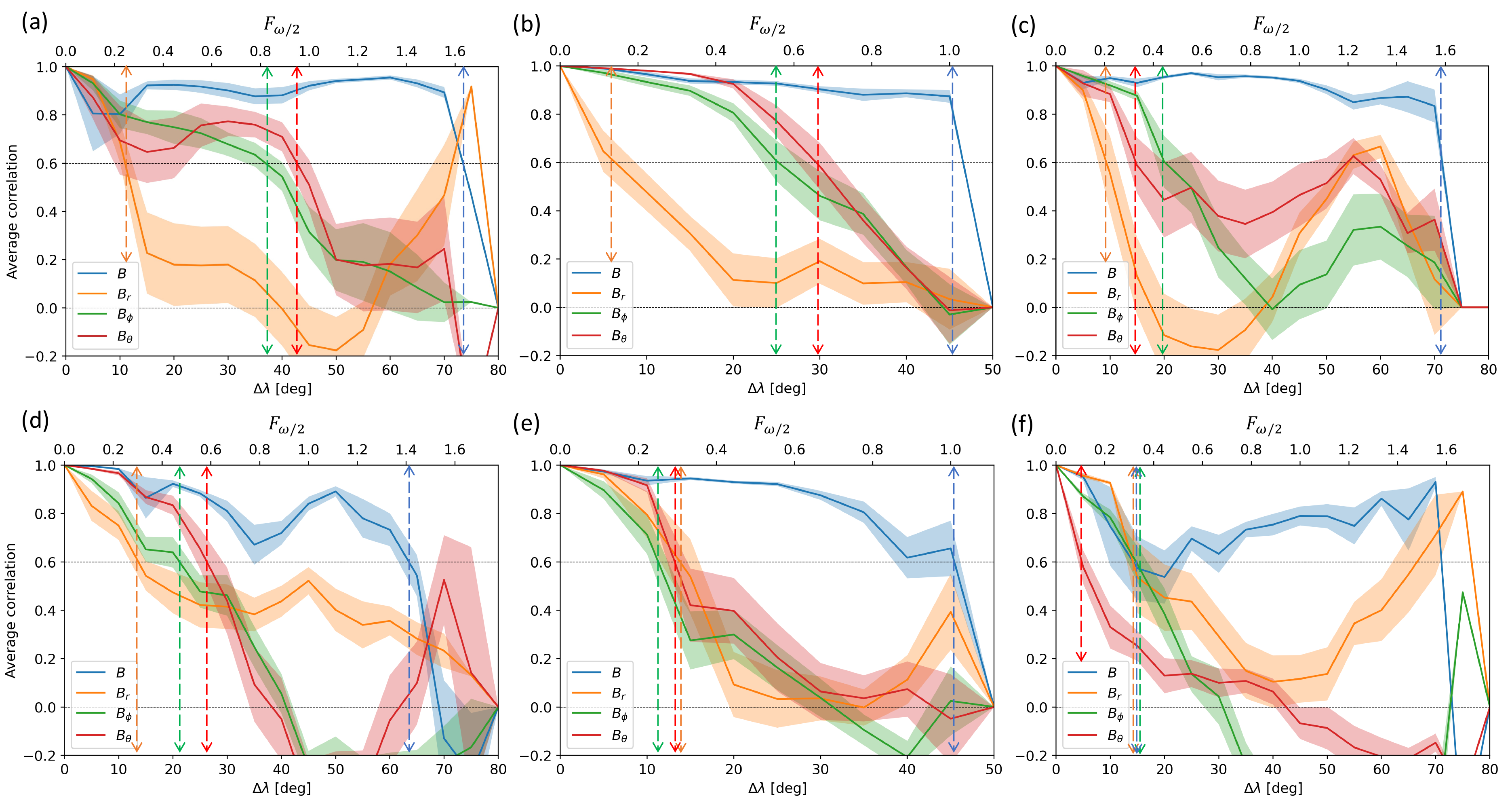}}
\caption{
Average correlation profiles for the magnetic field magnitude and components at 1~au, for run~A (top) and run~B (bottom) in EUHFORIA. 
Three reference {observers} are considered: one on the east (left), one near the center (center), and one on the west (right) of the {ME}.
{In all panels, the solid lines indicate the correlations calculated by averaging across all possible directions with respect to the reference observer, while the shaded areas} represent the error bars estimated from the standard errors, i.e.\ $\pm \mathrm{SE}_{\mathrm{corr}B}$, $\pm \mathrm{SE}_{\mathrm{corr}Br}$, $\pm \mathrm{SE}_{\mathrm{corr}B\theta}$, $\pm \mathrm{SE}_{\mathrm{corr}B\phi}$, obtained from considering different directions around the reference observer.
The average radii of correlation, measures of the characteristic correlation scales within the ME, are marked by the dashed arrows.
} 
\label{fig:euhforia_correlations}
\end{figure*}
%
Figure~\ref{fig:euhforia_correlations} shows the results of the analysis for the average correlation scale of the magnetic field magnitude and {the $B_r$, $B_\theta$ and $B_\phi$} components for run~A {(panels (a) to (c))} and B {(panels (d) to (f))} at 1~au, for {these three reference observers}.

{By comparing the results from runs~A (Figure~\ref{fig:euhforia_correlations} (a)--(c)) and~B (Figure~\ref{fig:euhforia_correlations} (d)--(f)), we find that all magnetic field components (except for $B_r$) have smaller average correlation scales in run~B than in run~A, indicating} interactions may significantly reduce the ability of {MEs} to behave as magnetically coherent structures. 
For an observer crossing near the center of the {ME}, $B$ remains highly correlated throughout the whole magnetic structure ($\Delta \lambda = 45^\circ$, {corresponding to $F_{\omega/2}=1.00$ of the initial {CME} half angular width)} for both runs A {(panel (b))} and B {(panel (e))}. These scales are slightly higher than those reported for an ideal spheromak structure ($\Delta \lambda = 40^\circ$, $F_{\omega/2}=0.88$, {not shown}). We speculate that expansion/deformation mechanisms (such as those reported within 0.3~au by \citet{Scolini2021a}; see Figure~12 therein) may be acting across the whole {ME} cross-section during early propagation phases (from 0.1 to 0.3~au) leveling out spatial asymmetries in the total magnetic field profile. 
{The} average correlation scales for $B_\theta$ and $B_\phi$ drop from $\Delta \lambda \sim 25^\circ-30^\circ$ ($F_{\omega/2} \sim 0.55-0.66$) in run~A, to $\Delta \lambda \sim 10^\circ-15^\circ$ ({$F_{\omega/2} \sim 0.22-0.33$}) in run~B. In run~A the correlation scale is already significantly reduced compared to the ideal spheromak case (where $\Delta \lambda = 40^\circ$, $F_{\omega/2}=0.88$, {not shown}), and even more so in run~B.
Only the average correlation scale for $B_r$ slightly increases from $\Delta \lambda = 5^\circ$ ($F_{\omega/2}=0.11$) to $\Delta \lambda = 15^\circ$ ($F_{\omega/2}=0.33$), and both runs report higher correlation scales than in the ideal spheromak case (where the correlation was lost only a few degrees away from the reference observer). 
However, $\mathrm{SE}_{\mathrm{corr}Br}$ is generally larger than for the other components in both EUHFORIA runs, reflecting the fact that the actual average correlation scale is highly asymmetric around the reference observer, and therefore is affected by large error bars.

{The results are qualitatively the same for an observer crossing through the east and west of the ME. 
For both observers, $B$ remains highly correlated throughout the whole magnetic structure in both runs~A and B ($\Delta \lambda = 65^\circ-70^\circ$, corresponding to $F_{\omega/2}=1.44-1.55$). 
The average correlation scales for $B_\theta$ and $B_\phi$ reduce by a factor $\sim2$ or more between run~A and run~B. 
For an east observer, the correlations reduce from $\Delta \lambda \sim 35^\circ-45^\circ$ ($F_{\omega/2} \sim 0.77-1.00$) in run~A to $\Delta \lambda \sim 20^\circ-25^\circ$ ({$F_{\omega/2} \sim  0.44-0.55$}) in run~B.
For a west observer, the average correlation scale for $B_\theta$ decreases from $\Delta \lambda = 15^\circ$ ($F_{\omega/2} = 0.33$) to $\Delta \lambda = 5^\circ$ ({$F_{\omega/2} = 0.11$}), while that of $B_\phi$ decreases from $\Delta \lambda = 20^\circ$ ($F_{\omega/2} = 0.44$) to $\Delta \lambda = 15^\circ$ ({$F_{\omega/2} = 0.33$}) between run~A and B. These results clearly demonstrate how correlations for a west observer in $B_\theta$ and $B_\phi$ are up to a factor 4 smaller than correlations for an east observer.
The average correlation scale for $B_r$ remains roughly constant around $\Delta \lambda \sim 10^\circ-15^\circ$ ($F_{\omega/2} \sim 0.22-0.33$) in both runs.}

{It is worth mentioning that} even in the case without interactions, smaller correlation scales are found for {a} west observer compared to {an} east observer. Due to the lack of structures in the simulated ambient solar wind in run~A, we consider this result to be linked to the propagation of the ICME through a magnetic field characterized by a Parker spiral configuration, which presents an intrinsic asymmetry with respect to the ICME propagation direction that leads to different ICME--solar wind interaction behaviors at the two ICME flanks \citep[e.g.\ in terms of pressure balance; see][]{Wang2004}. Differences between the east and west observers are even more pronounced in run~B, and this is interpreted as due to the superposition of the effects related to the east-west Parker spiral asymmetry, and the interaction with the HSS/SIR mainly affecting the western portion of the ICME. 

{In light of these results, we clarify why we use a different definition of correlation scale compared to observational works \citep{Lugaz2018, Wicks2010}. First of all, in both the ideal spheromak structure and EUHFORIA simulations we found no scale at which correlations in the $B$, $B_\theta$ and $B_\phi$ components reach values near zero. This is more likely due to the very idealized ME structure considered, while actual {MEs} may have more complex configurations already following their eruption in the low solar corona and before reaching 0.1~au. Secondly, contrary to the ME and solar wind profiles reported by \citet{Lugaz2018} and \citet{Wicks2010}, our correlation does not scale in a quasi-linear or quasi-exponential manner, so we decided to refrain from fitting curves with specific functions and rather defined a measure of coherence that was more appropriate for our specific problem.}

{Despite these methodological differences, the results retrieved in our numerical simulations have several elements in common with observational estimates. Specifically, 
t}he ranking of the average correlation scales for the different magnetic field components {(with $B$ and $B_r$ being the most and least correlated, respectively)} is consistent with what was found within an ideal spheromak structure and with estimates obtained from in situ observations of real {MEs} by \citet{Lugaz2018}. 
Furthermore, the range of average correlation scales for $B$ (estimated around $45^\circ$ for a central observer, and $15^\circ$ to $70^\circ$ for a flank observer) encompasses previous constraints on the characteristic scale of magnetic coherence at 1~au by \citet{Lugaz2018} ($12^\circ-20^\circ$) and \citet{Owens2020} ($\sim 26^\circ$). 
The range of average correlation scales for $B_\theta$ and $B_\phi$ (estimated around $10^\circ$ to $30^\circ$ for a central observer, and between $5^\circ$ and $45^\circ$ for a flank observer) also encompasses previous estimates on the characteristic scale of magnetic coherence at 1~au for the magnetic field components, reported in range $4^\circ-7^\circ$ by \citet{Lugaz2018}. 
The larger upper limit found in our study is somewhat expected: the global, ideal MHD nature of our numerical model, as well as the simplified set-up used in this study (including the insertion of a perfectly-known global {CME} magnetic structure at the inner boundary of the heliospheric domain, located at 0.1~au in our simulations) likely resulted in a more regular/smooth magnetic configuration in simulated {MEs} compared to real {MEs} at the same heliocentric distance. Additionally, we note the diverse nature of previous studies (in terms, e.g., of working assumptions, data sets, range of parameters explored) when compared to each other and to this investigation, which may have also led to different coherence scale estimates across these studies.
 
\begin{figure*}
\centering
{\includegraphics[width=\hsize]{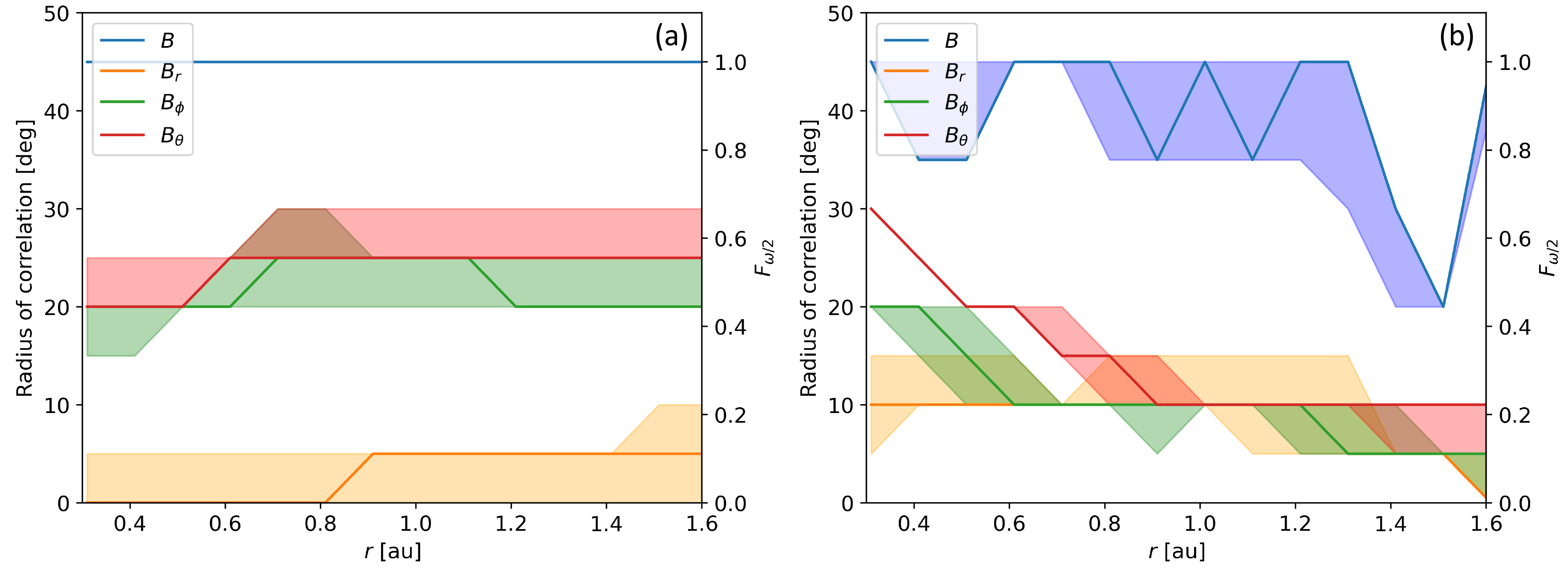}}
\caption{
Characteristic scale of correlation for the total magnetic field and the three magnetic field components, as a function of the heliocentric distance, for the two spheromak CMEs simulated in EUHFORIA. The reference location is taken at $(\theta, \phi)=(0^\circ,0^\circ)$ for run~A (a), and $(\theta, \phi)=(5^\circ,0^\circ)$ for run~B (b), corresponding to the CME initial direction in the two simulations.
{In all panels, the continuous lines indicate the average correlation scale with respect to the reference observer, while the shaded areas} represent the error bars {on the average correlation scale} estimated from the standard errors, i.e.\ $\pm \mathrm{SE}_{\mathrm{corr}B}$, $\pm \mathrm{SE}_{\mathrm{corr}Br}$, $\pm \mathrm{SE}_{\mathrm{corr}B\theta}$, $\pm \mathrm{SE}_{\mathrm{corr}B\phi}$, obtained from considering different directions around the reference observer.
} 
\label{fig:euhforia_correlation_scale}
\end{figure*}

We conclude the analysis of magnetic coherence by considering the evolution of the {ME} average correlation scales with heliocentric distance. {In Figure~\ref{fig:euhforia_correlation_scale}, we show the ME} average correlation scales for the total magnetic field and the three magnetic field components, as a function of the heliocentric distance, for the two EUHFORIA simulations and a reference observer crossing near the {ME} center. {We find that the} characteristic correlation scales for an ICME propagating through a homogeneous solar wind are unchanged during its propagation {(Figure~\ref{fig:euhforia_correlation_scale}~(a))}, remaining around $F_{\omega/2}=1$ for $B$, $F_{\omega/2}=0.44-0.66$ for $B_\theta$ and $B_\phi$, and below $F_{\omega/2}=0.22$ for $B_r$ {at all heliocentric distances}. On the other hand, an interaction with a HSS/SIR induces a clear decrease in the {ME} average correlation scales during propagation, particularly for $B_\theta$ and $B_\phi$ (Figure~\ref{fig:euhforia_correlation_scale}~(b)). 
The decrease is more rapid between 0.3~au and 1~au, while it continues at smaller rates up to 1.6~au. 
Overall, the maximum scale of correlation reduces from $F_{\omega/2} = 1$ to $0.44$ for $B$, from $F_{\omega/2} = 0.66$ to 0.11 for $B_\theta$, from $F_{\omega/2} = 0.44$ to 0 for $B_\phi$, and from $F_{\omega/2} = 0.33$ to 0 for $B_r$ between 0.3~au and 1.6~au.
Interactions with other large-scale structures may therefore induce a decrease by a factor of 3 to 6 in the {ME} magnetic field correlation scales.

\section{Implications for Ecliptic Missions}
\label{sec:discussion}

\begin{figure*}
\centering
{\includegraphics[width=\hsize]{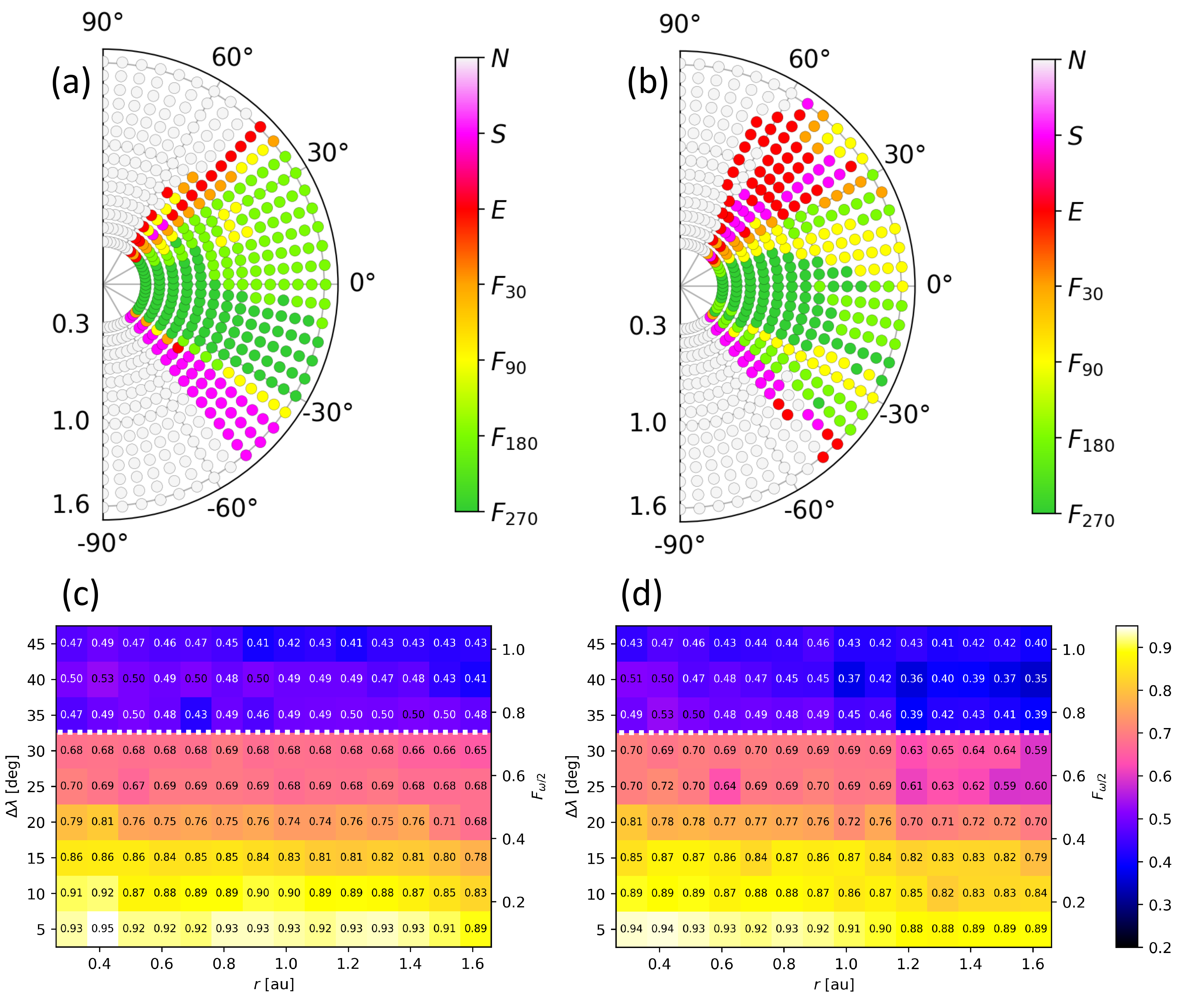}}
\caption{
Analysis of the ME type distribution and Moran's $I$ as a function of the heliocentric distance, at $\theta = \theta_\mathrm{CME}$, for the two ICMEs simulated in EUHFORIA.
(a), (b): distribution of ME types {(color coded) as a function of the heliocentric distance and longitude} for runs A and B, respectively.
(c), (d): Moran's $I$ {(also color coded)} as a function of the {heliocentric distance and} spacecraft separation $\Delta \lambda$, for runs A and B, respectively.
{The dotted white lines distinguish between regions where $I > 0.5$ (considered regions with well-ordered distribution of ME types and low magnetic complexity) and regions where $I \le 0.5$ (considered regions with randomly-ordered distribution of ME types and high magnetic complexity).}
} 
\label{fig:euhforia_equator}
\end{figure*}

{In this section, we focus on spacecraft near the CME equatorial plane (which is close to the solar equatorial and ecliptic planes in our particular case) in order to support the planning and interpretation of observations by future multi-spacecraft missions exploring ICME structures in the ecliptic plane. The simulation set-up considered here specifically applies to spacecraft in the near-ecliptic region that cross through low-inclination FRs, which constitute about half of the ME FR configurations detected within 1~au \citep[e.g.][]{Lepping1990, Bothmer1998, Lynch2003}, and the majority of cases observed between 1~au and 2.2~au \citep{Davies2022}.}

{In order to determine the main implications for ecliptic missions, we focus on spacecraft crossings in our} EUHFORIA-simulated ICMEs close to the ecliptic plane and near the spheromak central magnetic axis, i.e.\ at $\theta = \theta_\mathrm{CME}$. $\theta_\mathrm{CME}$ corresponds to $\theta = 0^\circ$ in run~A and to $\theta = 5^\circ$ in run~B, and we chose these latitudinal values to more fairly compare crossings across the ICMEs between the two runs.
{The main results are displayed in Figure~\ref{fig:euhforia_equator}~(a) and (b), which show the spatial distribution of ME types detected at different heliocentric distances by simulated spacecraft located at $\theta = \theta_\mathrm{CME}$ in runs A and B.} 
{In the case of run A, i.e. ICME with no interaction, Figure~\ref{fig:euhforia_equator}~(a)} shows a relatively well-ordered distribution of ME types, with spacecraft aligned in different radial directions recording similar ME types as the {ME} propagates to larger heliocentric distances. The dominance of $F_{270}$ and $F_{180}$ types (associated with {$\mathbb{C}=0$ and $1$, respectively}) suggest{s} the {ME} is less complex in its equatorial region than it is globally. {This result is consistent with the spatial distribution of ME types obtained for an ideal spheromak structure (as shown in Figure~6~(d) in \citet{Scolini2021b}).} Interestingly, we find that even without interactions with large-scale structures, a change in the most frequently observed ME type is observed around 0.8~au, where $F_{270}$ types are taken over by $F_{180}$. $F_{270}$ types are initially dominant and spread across different directions, but progressively concentrate only in the eastern part of the {ME}, while $F_{180}$ types become dominant in the western part. As discussed above, this is likely a reflection of Parker spiral effects induced by propagation through an asymmetric interplanetary magnetic field. $F_{90}$, $F_{30}$, and $E$ types {(corresponding to $\mathbb{C}=2,3$ and $4$, respectively)} are all very rarely observed in the ecliptic plane, making up less than 20\% of the observed cases.
The analysis of the results considering different angular separations (not shown) reveals that only spacecraft separated by $\Delta \lambda \lesssim 15^\circ$ ($F_{\omega/2} \lesssim 0.33$, corresponding to $\sim6$ spacecraft crossing the {ME} in different directions near its central axis) {can characterize} the ecliptic complexity of the {ME} {observed by} spacecraft {at} $\Delta \lambda \lesssim 5^\circ$ ($F_{\omega/2} \lesssim 0.11$) across all heliocentric distances{. However, we emphasize that regardless of their angular separation, observations in the ecliptic plane alone are not sufficient to characterize the global complexity of the {ME} (where $F_{270}$ and $F_{30}$ were the two dominant ME types; see Section~\ref{subsec:euhforia_me_types}), and off-ecliptic observations are therefore required.}

{In run B (ICME with interaction), we observe a much more irregular distribution of the ME types in both the longitudinal and radial directions (Figure~\ref{fig:euhforia_equator}~(b)).}  {On the ME} equatorial plane we find that the most common ME types are $F_{270}$ up until 1~au, {after which} $F_{180}$, $F_{90}$ and $E$ types become comparable. Just as for run~A, we therefore conclude that the {ME} is less complex near its central axis than it is globally. Results at different angular separations also indicate that a swarm of at least $\gtrsim 9$ near-equatorial spacecraft separated by $\Delta \lambda \lesssim 10^\circ$ ($F_{\omega/2} \lesssim 0.22$) would be required to capture the ecliptic {ME} spatial complexity and its evolution across all heliocentric distances. {Furthermore, similarly to} the case of run~A, off-ecliptic observations are required to characterize the global complexity of the {ME as} discussed in Section~\ref{subsec:euhforia_me_types}.

{To calculate the Moran's $I$, we have to consider a 2-D spacecraft array distributed in longitude and latitude (see Equation~\ref{eqn:moransI}). Therefore, we compute the Moran's $I$ at each heliocentric distance from near-equatorial regions in EUHFORIA simulations by considering a near-equatorial array of spacecraft within $\pm 15^\circ$ from $\theta=\theta_\mathrm{CME}$. This array consists of $\Delta \phi = \Delta \theta = 5^\circ$ separated spacecraft arranged in $3\times2+1 = 7$ latitudinal rows.}
Figure~\ref{fig:euhforia_equator}~(c) and (d) report the results of the analysis. In general, we observe higher values of $I(r)$ compared to Figure~\ref{fig:euhforia_moransI}, indicating that the {ME} exhibits a higher spatial autocorrelation in near-equatorial regions than it does in the perpendicular direction. {We obtain a similar result in the case of an ideal spheromak structure (not shown), suggesting that an ideal or weakly-perturbed spheromak structure is particularly ordered near its magnetic axis.}
For run~A (Figure~\ref{fig:euhforia_equator}~(c)), we find that at {all heliocentric distances}, $I(r)$ decreases with $\Delta \lambda$ and remains above 0.5 up until $\Delta \lambda=30^\circ$ ($F_{\omega/2}=0.66$), which is slightly larger than the separation scale obtained when considering the whole {ME} structure (Figure~\ref{fig:euhforia_moransI}~(a)). This further confirms the higher spatial clustering of ME types within the {ME} in the near-equatorial regions compared to other directions. Results are also slightly decreasing with heliocentric distance at scales $\Delta \lambda \le 20^\circ$ ($F_{\omega/2}=0.44$), while they are independent on the propagation phase at larger angular separations.
For run~B (Figure~\ref{fig:euhforia_equator}~(d)), we obtain similar results to run~A in terms of the scale at which $I(r)$ becomes smaller than 0.5 ($\Delta \lambda=30^\circ$, corresponding to $F_{\omega/2}=0.66$). However, {we also notice that $I(r)$ shows} a more clear dependence on the heliocentric distance up to scales $\Delta \lambda \le 40^\circ$ ($F_{\omega/2}=0.88$), suggesting that the perturbation to the {ME} induced by the interaction with the HSS/SIR extends across a wider range of angular scales in the near-equatorial plane compared to other directions (Figure~\ref{fig:euhforia_moransI}~(b)).

\begin{figure*}
\centering
{\includegraphics[width=\hsize]{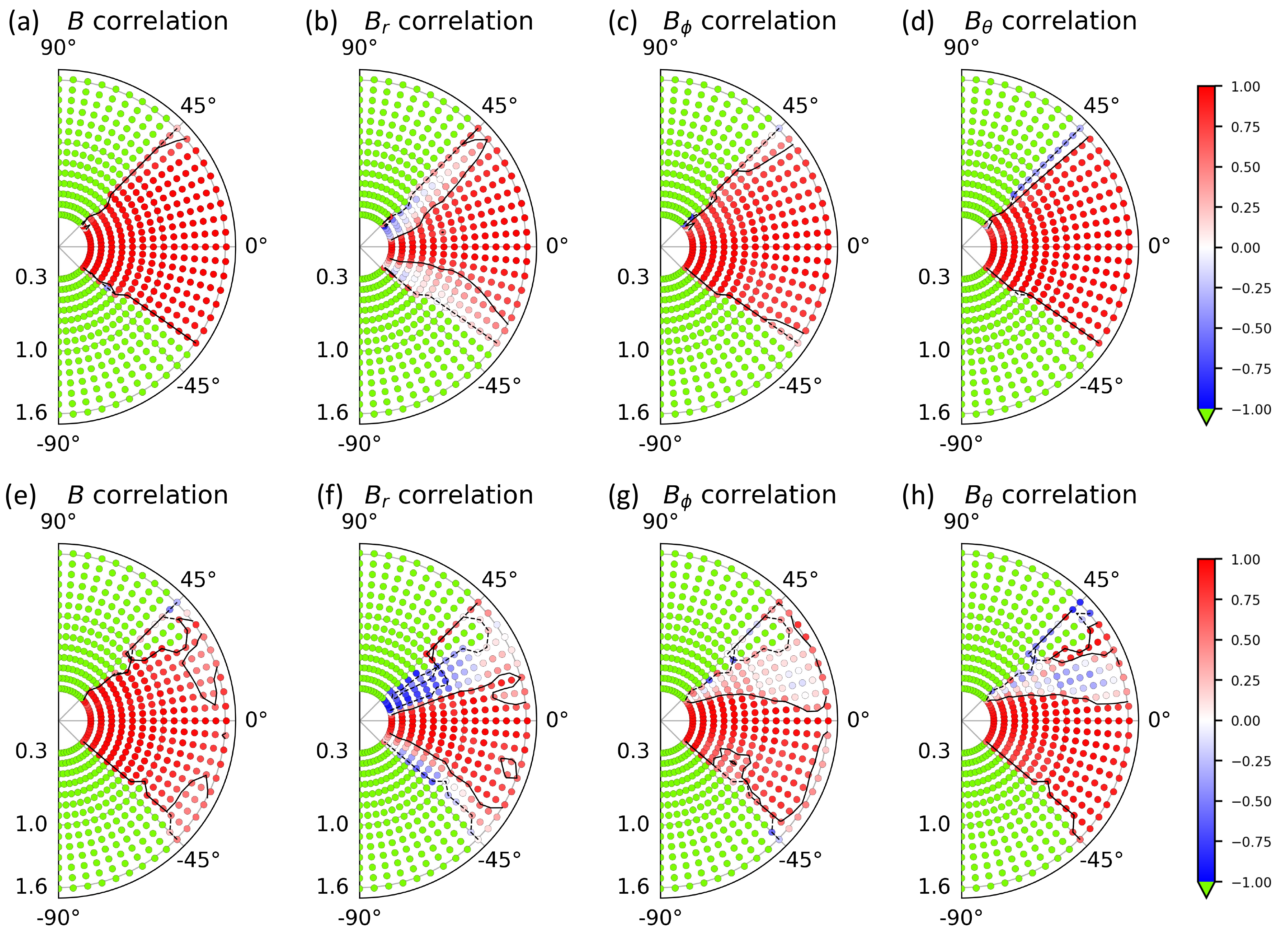}}
\caption{
Correlation maps {(color coded)} for the total magnetic field and the three magnetic field components with respect to a central reference observer {($\phi_\mathrm{ref}=0^\circ$), as a function of the heliocentric distance and longitude} at $\theta = \theta_\mathrm{CME}$, for run~A (top) and run~B (bottom) in EUHFORIA. 
(a), (e): $B$.  
(b), (f): $B_r$.
(c), (g): $B_\phi$.
(d), (h): $B_\theta$.
{In each panel, b}lack contours mark regions were the correlation is $+0.6$ (continuous lines) or $-0.6$ (dashed lines).
} 
\label{fig:euhforia_equator_correlation}
\end{figure*}

The correlation with respect to a central reference observer tends to be maximum along the {ME} magnetic axis (i.e.\ near the equatorial plane){: focusing on the results at $\theta = \theta_\mathrm{CME}$, we find that for both an ideal spheromak structure (not shown) and for run~A (as shown in Figure~\ref{fig:euhforia_equator_correlation}~(a) to (d))}, the correlation extends throughout the whole {ME} width. {These results confirm that an ideal or weakly-perturbed spheromak structure exhibits coherent signatures across wider scales particularly along its principal magnetic axis.}
In run~B (panels (e) to (h)), the correlation scale remains generally large for $B$ up to 1.3~au, and it then reaches a minimum of $5^\circ$ ($F_{\omega/2} = 0.11$) in the west direction {and} $20^\circ$ ($F_{\omega/2} = 0.44$) in the east direction around 1.5~au. 
For $B_\theta$ and $B_\phi$, the {ME} westernmost region {becomes} progressively more perturbed than the easternmost one during propagation, up to the point of losing {all correlations} with a central reference observer. Near 1.5~au, the correlation scale is between $\lesssim 5^\circ$ ($F_{\omega/2} \lesssim 0.11$) in the west direction and $35^\circ$ ($F_{\omega/2} = 0.77$) in the east direction for $B_\phi$; and between $5^\circ$ ($F_{\omega/2} = 0.11$) in the west direction and $45^\circ$ ($F_{\omega/2} = 1$) in the east direction for $B_\theta$.

{Finally, we note that for spacecraft crossings near the spheromak magnetic field axis (either for an {ideal} spheromak structure, or in EUHFORIA simulations), ME signatures are expected to be similar to the ones encountered by crossing near the axis of cylindrical FR geometries, i.e.\ with bipolar $B_\theta$ signatures and nearly-unipolar signatures in $B_\phi$ \citep[see Figure~6 in][]{Scolini2021b}. Results along the direction $\theta = \theta_\mathrm{CME}$ in each EUHFORIA simulation can therefore be expected to be less dependent on the specific FR model used than in other directions, {and we hypothesize they could be potentially extrapolated to other FR models as well}. For the opposite reason, the most severe difference between the spheromak model and other cylindrical FR models is expected to be found in the direction perpendicular to the magnetic axis (e.g.\ as in the case of ecliptic crossings through high-inclination FR types), but we leave the consideration of this alternative scenario for future studies.}

\section{Conclusions}
\label{sec:conclusions}

\begin{table}
\footnotesize
\centering
    \begin{tabular}{l||cc|cc}
         & \multicolumn{2}{c|}{Whole structure} & \multicolumn{2}{c}{Equatorial plane}\\
         & {Run~A} & {Run~B} & {Run~A} & {Run~B}\\
         \hline\hline
        Complexity: ME types                    & {$25^\circ$}(0.55) & {$10^\circ$}(0.22)   & {$15^\circ$}(0.33)    & {$10^\circ$}(0.22)  \\
        Complexity: autocorrelation     & {$25^\circ$}(0.55) & {$20^\circ$}(0.44)   & {$30^\circ$}(0.66)    & {$30^\circ$}(0.66) \\
        Coherence: $B$                          & 45$^\circ$(1.00)     & 20--45$^\circ$(0.44--1.00) & 35--40$^\circ$(0.77--0.88)  & 5--20$^\circ$(0.11--0.44) \\
        Coherence: $B_r$                        & 0--10$^\circ$(0.00--0.22)  & 0--5$^\circ$(0.00--0.11)  & 10--25$^\circ$(0.22--0.55)  & 5--10$^\circ$(0.11--0.22) \\
        Coherence: $B_\theta$                   & 15--30$^\circ$(0.33--0.66) & 0--5$^\circ$(0.00--0.11)  & 35--40$^\circ$(0.77--0.88)  & 5--45$^\circ$(0.11--1.00) \\
        Coherence: $B_\phi$                     & 20--30$^\circ$(0.44--0.66) & 5--10$^\circ$(0.11--0.22) & 30--35$^\circ$(0.66--0.77)  & 5--35$^\circ$(0.11--0.77) \\
    \end{tabular}
    \caption{
    Summary of the characteristic scales of complexity and coherence within {MEs} for a central observer (between 0.3 and 1.6~au) retrieved in this work. 
    Values are expressed in terms of the spacecraft angular separation $\Delta \lambda$ and of the fraction of CME initial half {angular} width $F_{\omega/2}$ (in brackets).}
    \label{tab:results}
\end{table}

In this study, we presented an in-depth investigation of the spatial scales of magnetic complexity and coherence within {MEs} as inferred from spacecraft swarms in global heliospheric simulations based on the consideration of various ICME-solar wind interaction scenarios and different complexity and coherence metrics. This work is also the first-ever attempt to investigate {ME} magnetic coherence scales by analyzing the results of global numerical simulations in an observationally-consistent fashion. 
{Specifically, we evaluated the magnetic complexity of the simulated MEs based on the frequency of ME types detected by virtual spacecraft crossing the MEs along different directions during their propagation through interplanetary space. We further explored the spatial distribution of magnetic complexity within the simulated MEs using the Moran's $I$ spatial autocorrelation index. We determined the characteristic scale of magnetic coherence within the MEs by evaluating the maximum angular separation at which virtual spacecraft crossing the same ME observed magnetic field signatures highly correlated to those detected by a reference observer.}
A summary of the characteristic scales derived from numerical simulation between 0.3 and 1.6~au is provided in Table~\ref{tab:results}, and the main conclusions are summarized as follows:
\begin{enumerate}
\item On the one hand, analyses of the frequency and spatial clustering of ME types within {MEs} indicate that their global magnetic complexity does not change with heliocentric distance if no interaction with other large-scale solar wind structures occurs during their propagation. On the other hand, the {ME} magnetic complexity is likely to increase during its propagation if interactions with other large-scale solar wind structures occur, confirming previous results by \citet{Winslow2016, Winslow2021a} and \citet{Scolini2021b, Scolini2022}.
\item The two global magnetic complexity metrics considered (frequency of ME types, and {the Moran's $I$ spatial autocorrelation index}) both indicate that the consideration of smaller angular scales is required to capture global complexity changes in the presence of interactions. {As demonstrated in Section~\ref{subsec:euhforia_me_types}, f}or an ICME {with initial half} angular width of ${45}^\circ$ as the one simulated in this work, the spatial characterization of the occurrence of ME types requires in situ observations at separations no larger than $\Delta \lambda = 25^\circ$, but separations no larger than $\Delta \lambda = 10^\circ$ are needed in the case of ICME--solar wind interacting scenarios. {Section~\ref{subsec:euhforia_moransI} proved that} the analysis of the spatial autocorrelation of ME types via the Moran's $I$ coefficient requires observations at separations up to $\Delta \lambda = 25^\circ$ for a non-interacting ICME, which reduces to $\Delta \lambda = 20^\circ$ or less for ICME--HSS/SIR interacting scenarios.
{The angular separations above are more conveniently expressed in terms of scale-invariant quantities, particularly as fractions of the CME initial half {angular} width ($F_{\omega/2}$) and in terms of the associated number of virtual spacecraft required to cross the {ME} structure, which are independent of the prescribed {initial} CME half {angular} width and can therefore be extrapolated to CMEs of different sizes. In this respect, we found that the spatial characterization of the occurrence of ME types within an {ME} requires in situ observations at separations $F_{\omega/2} \le 0.55$, corresponding to a minimum of 7 to 12 spacecraft crossings. Separations $F_{\omega/2} \le 0.22$, corresponding to a minimum of 50 to 65 spacecraft crossings, are needed in the case of ICME--solar wind interacting scenarios. 
The analysis of the spatial autocorrelation of ME types via the Moran's $I$ coefficient requires observations at separations up to $F_{\omega/2} = 0.55$, corresponding to 7 to 12 spacecraft crossings for a non-interacting ICME, which reduces to $F_{\omega/2} \le 0.44$, corresponding to 13 to 16 spacecraft crossings or less for ICME--HSS/SIR interacting scenarios.}
\item In our numerical simulations, the scale of correlation of the magnetic field components within MEs {was different} for different components, and tended to be the largest for $B$, followed by $B_\theta$ and $B_\phi$, and ultimately by $B_r$. {As discussed in Section~\ref{subsec:euhforia_magnetic_coherence}, the} characteristic scale of correlation {reduced} by a factor of 3--6 {for MEs within} ICMEs interacting with preceding HSSs/SIRs ($\Delta \lambda = 5^\circ$, {corresponding to $F_{\omega/2} = 0.11$}) compared to ICMEs that do not interact with other large-scale structures ($\Delta \lambda = 15^\circ-30^\circ$, {corresponding to $F_{\omega/2} = 0.33-0.66$}). Results at different heliocentric distances also indicated that as ICMEs propagate to larger heliocentric distances, intervening interactions may efficiently destroy any pre-existing correlation scale {within their MEs}.
{Overall, such results suggest that the scale of magnetic coherence of {ME} structures would be optimally investigated by having spacecraft crossing through them at angular separations between $\Delta \lambda = 15^\circ$ and $30^\circ$ ($F_{\omega/2}=0.33-0.66$) in cases where no significant interactions with other large-scale interplanetary structures occur. However, angular separations down to $\Delta \lambda = 5^\circ$ ($F_{\omega/2}=0.11$) and below, are needed in cases where significant interactions with other large-scale interplanetary structures do occur. }

{Similar conclusions were reached by \citet{Farrugia2011} when analyzing an ICME interacting with a following HSS/SIR and observed at 1~au by three spacecraft at $\sim20^\circ$ separation. At the time, the authors emphasized the need for numerical simulations and/or more in situ observations for a complete elucidation of {ICMEs} interacting with SIRs, and the present work helps identify the most appropriate scales to be targeted in future studies to fill in current observation and knowledge gaps.
Additionally, our results indicate that achieving small spacecraft angular separations would be particularly important at larger distances from the Sun, as the correlation scale is found to progressively shrink during propagation, the longer the interaction with large-scale solar wind structures occurs.
For the opposite reason, when considering multi-spacecraft observations obtained in radial alignment, making sure that the angular offset among the different spacecraft involved is smaller than the aforementioned expected scales of coherence would be most appropriate to distinguish changes in magnetic signatures that are actually due to evolutionary effects, rather than to the intrinsic lack of correlation affecting observations acquired at larger angular separation scales.}
\item At least for fast ICMEs as the ones simulated in this work, the magnetic complexity inferred from the spatial autocorrelation of ME types via the Local Moran's $I$ coefficient was found to be higher in the western portion of MEs than on the eastern one regardless of the presence or absence of interactions with other large-scale solar wind structures. Additionally, smaller correlation scales were found around a reference observer crossing on the western portion of the {ME}, compared to one on crossing the eastern portion. East-west differences were more prominent when an interaction with a preceding HSS/SIR was involved, but were present also in the ICME propagating through a quasi-homogeneous solar wind. These results are found to be due to a combination of two effects: a Parker spiral contribution that affects both interacting and non-interacting ICMEs; and a contribution from the actual interaction of ICMEs with preceding large-scale solar wind structures, which is localized in the region most influenced by the interaction.
\item Analyses of the magnetic complexity and coherence in the near-ecliptic plane (approximately corresponding to the direction of the {ME} magnetic axis) {in Section~\ref{sec:discussion} indicated} that {MEs} may exhibit a lower complexity, higher spatial autocorrelation, and higher coherence along their magnetic axis compared to perpendicular directions. In this case, a minimum of $\sim6$ spacecraft crossings separated by $F_{\omega/2} \lesssim 0.33$ (corresponding to $\Delta \lambda \lesssim 15^\circ$) is necessary to characterize the evolution of {ME} magnetic complexity in the near-equatorial plane, which increases to $\sim9$ spacecraft ($\Delta \lambda \lesssim 10^\circ$, corresponding to $F_{\omega/2} \lesssim 0.22$) when interaction with preceding HSSs/SIRs are involved. However, also in the case of interactions with preceding large-scale solar wind structures, near-ecliptic observations alone were not able to characterize the {ME} global, off-ecliptic complexity.

\end{enumerate}

Overall, our estimates for the scale of magnetic correlation of $B_\theta$ and $B_\phi$ encompass previous constraints on the characteristic scale of magnetic coherence at 1~au, reported in range $4^\circ-7^\circ$ by \citet{Lugaz2018}. The range of correlation scales obtained for $B$ is also consistent with previous estimates by \citet{Lugaz2018} ($12^\circ-20^\circ$) and \citet{Owens2020} ($\sim 26^\circ$). However, due to the numerical model and set-up's specificities, and the multi-scale nature of the magnetic coherence problem, larger upper limits found in our study are somewhat expected. Additionally, while from an observational standpoint the correlation of magnetic field profiles among different locations within an {ME} has been regarded as a proxy for magnetic coherence \citep{Lugaz2018}, it also remains unclear how the ability of information to propagate across an {ME} affects its local properties, and how this relates to the ability of the same {ME} to use such information in order to resist deformation by external factors and behave as a coherent structure \citep{Owens2020}. While this work helps constrain some of the critical scales controlling the evolution of {ME} structures, and establishes new pathways for the investigation of {ME} magnetic coherence by analyzing the results of global numerical simulations in an observationally-consistent fashion (i.e.\ using spacecraft swarms), in the future, we envisage more sophisticated numerical tools in combination with new data from recently-launched missions (e.g.\ Parker Solar Probe and Solar Orbiter) will help pinpoint the scale and meaning of magnetic coherence within ICMEs more precisely.

\acknowledgments 
C.S. acknowledges the NASA Living With a Star Jack Eddy Postdoctoral Fellowship Program, administered by UCAR's Cooperative Programs for the Advancement of Earth System Science (CPAESS) under award no.\ NNX16AK22G. 
R.M.W. acknowledges support from NASA grant 80NSSC19K0914 as well as partial support from the NASA STEREO Transition grant.
N.L. was supported by NASA grants 80NSSC20K0700 and 80NSSC19K0831.
S.P.\ acknowledges support from the European Union's Horizon 2020 research and innovation programme under grant agreement No 870405 (EUHFORIA 2.0), the ESA project ``Heliospheric modelling techniques'' (Contract No. 4000133080/20/NL/CRS), and the projects C14/19/089 (C1 project Internal Funds KU Leuven), G.0D07.19N (FWO-Vlaanderen), SIDC Data Exploitation (ESA Prodex-12), and B2/191/P1/SWiM (Belspo).
EUHFORIA is developed as a joint effort between the University of Helsinki and KU Leuven.
Simulations were carried out at KU Leuven and the VSC -- Flemish Supercomputer Center (funded by the Hercules foundation and the Flemish Government -- Department EWI).

\bibliography{Refs}{}
\bibliographystyle{aasjournal}

\end{document}